\documentclass[12pt, draftclsnofoot, onecolumn]{IEEEtran}
\usepackage{amssymb}
\usepackage{amsmath}
\usepackage{cite}
\usepackage{url}
\usepackage{xcolor}
\usepackage{cite,graphicx,amsmath,amssymb}
\usepackage{subfigure}
\usepackage{citesort}
\usepackage{fancyhdr}
\usepackage{mdwmath}
\usepackage{mdwtab}
\usepackage{caption}
\usepackage{amsthm}
\usepackage{setspace}
\usepackage{multirow}
\usepackage{epstopdf}
\usepackage[justification=centering]{caption}
\newtheorem{remark}{Remark}
\newtheorem{theorem}{Theorem}

\newtheorem{lemma}{Lemma}

\newtheorem{corollary}{Corollary}

\newtheorem{proposition}{Proposition}

\usepackage{algorithm}
\usepackage{algorithmic}
 \usepackage{cases}

\begin{document}
	
	\title{Reconfigurable Intelligent Surface Assisted Cooperative Non-orthogonal Multiple Access Systems}
	\author{
		Jiakuo~Zuo,
		Yuanwei~Liu,~\IEEEmembership{Senior Member,~IEEE,}
		and Naofal Al-Dhahir,~\IEEEmembership{Fellow,~IEEE}
		\thanks{J. Zuo is is with the School of Internet of Things,  Nanjing University of Posts and Telecommunications, Nanjing 210003, China (e-mail: zuojiakuo@njupt.edu.cn).}
		\thanks{Y. Liu is with the School of Electronic Engineering and Computer Science, the Queen Mary University of London, London E1 4NS, U.K.  (e-mail: yuanwei.liu@qmul.ac.uk).}
		\thanks{N. Al-Dhahir is with the Department of Electrical and Computer Engineering, The University of Texas at Dallas, Richardson,
		TX 75080 USA. (email:aldhahir@utdallas.edu).}
}
	\maketitle
	\vspace{-1cm}
	\begin{abstract}
		This paper considers downlink of reconfigurable intelligent surface (RIS) assisted cooperative non-orthogonal multiple access (CNOMA) systems. Our objective is to minimize the total transmit power by jointly optimizing the active beamforming vectors, transmit-relaying power, and RIS phase shifts. The formulated problem is a mixed-integer nonlinear programming (MINLP) problem. To tackle this problem, the alternating optimization approach is utilized to decouple the variables. In each alternative procedure, the optimal solutions for the active beamforming vectors, transmit-relaying power and phase shifts are obtained. However, the proposed algorithm has high complexity since the optimal phase shifts are solved by integer linear programming (ILP) whose computational complexity is exponential in the number of variables. To strike a good computational complexity-optimality trade-off, a low-complexity suboptimal algorithm is proposed by invoking the iterative penalty function based semidefinite programming (SDP) and the successive refinement approaches. Numerical results illustrate that:
		i) the proposed RIS-CNOMA system, aided by our proposed algorithms,  outperforms the conventional CNOMA system. ii) the proposed low-complexity suboptimal algorithm can achieve the near-optimal performance. iii) whether the RIS-CNOMA system outperforms the RIS assisted non-orthogonal multiple access (RIS-NOMA) system depends not only on the users' locations but also on the RIS' location.
	\end{abstract}
	\begin{IEEEkeywords}
		{B}eamforming, cooperation communication, half-duplex, non-orthogonal multiple access, reconfigurable intelligent surface.
	\end{IEEEkeywords}
	\section{Introduction}
	Recently, reconfigurable intelligent surface (RIS) technology has attracted significant attention since it is capable of providing cost-efficient and power-efficient high data-rate services for next generation communication systems~\cite{liu2020reconfigurable}. A RIS is a metasurface composed of a large number of low-cost passive elements. By properly configuring the phase shift and/or amplitude of each passive element, the received signal power can be boosted by constructively adding the reflected signals at the receiver. Different from the traditional amplify-and-forward (AF) relay and decode-and-forward (DF) relay, the RIS does not have signal processing capability. Instead, it only reflects the signals in a passive way~\cite{zhao2019survey}. Therefore, the RIS assisted wireless networks have lower hardware cost and energy consumption. With the advantages of small size and low cost, RISs can be flexibly deployed on building facades, road signs, lamp posts, etc. Furthermore, RISs can be integrated into existing communication systems with no change in the hardware at the base stations and access points as well as user terminals~\cite{wu2019towards,basar2019wireless}.

As a promising technique for supporting massive connectivity and enhancing spectrum efficiency, non-orthogonal multiple access (NOMA) has also received a considerable attention~\cite{9154358}. To further improve the performance of NOMA systems, an appealing extension for NOMA systms, namely, cooperative NOMA (CNOMA) systems, was proposed in~\cite{ding2015cooperative}. The key idea of CNOMA is to rely on the NOMA-strong users acting as DF relays to assist the NOMA-weak users.
By adding the cooperation transmission link as a new degree of diversity to the direct link transmission, CNOMA systems can provide spatial diversity to mitigate fading and extend coverage~\cite{dinh2020low}. In addition, the reliability of NOMA-weak user can be significantly improved by CNOMA systems. Thus, the fairness of NOMA transmission can be improved, particularly when the NOMA-weak user is at the edge of the cell~\cite{liu2018non}.

\subsection{Related Works}
\subsubsection{Studies on CNOMA Systems}
The research paradigms of CNOMA systems have been extensively studied under different application scenarios and objectives. For example, the performance of the downlink CNOMA system was studied in~\cite{dinh2020low}, where users have the capability of either half-duplex (HD) or full-duplex (FD) communications. To maximize the achievable sum rate, a two-step policy was proposed by jointly optimizing the user pairing and power control.
The two-way CNOMA system was first studied in~\cite{bae2019joint} and the exact closed-form outage probability expression was derived. To minimize the system outage probability, a joint power and time allocation algorithm was proposed.
The CNOMA system in multiple-input-multiple-output (MIMO) channels was studied in~\cite{li2018cooperative}, where the objective was to maximize the achievable rate by jointly optimizing the beamforming vectors of the base station and central user. A constrained convex-concave algorithm and a closed-form search based suboptimal algorithm were proposed to solve the problem.
The application of simultaneous wireless information and power transfer (SWIPT) to CNOMA system was investigated in~\cite{xu2017joint}. A new multiple-input single-output (MISO) SWIPT CNOMA system protocol was proposed by jointly optimizing the power splitting and beamforming vectors.
In~\cite{9088210}, the physical layer security problem was studied for the CNOMA system. Two power allocation schemes for the legitimate and jamming signals were proposed based on different knowledge assumptions of the eavesdropping channel state information.
\subsubsection{Studies on RIS Assisted NOMA Systems}
Some initial efforts to enhance the performance of RIS assisted NOMA (RIS-NOMA) systems with diverse optimization goals has been studied. In particular, for the single-input single-output (SISO) RIS-NOMA system in~\cite{zheng2020intelligent}, the transmit power minimization problem was formulated and the theoretical performance comparison between the RIS-NOMA system and RIS assisted orthogonal multiple access (RIS-OMA) system was investigated. 
In~\cite{ni2020resource}, the user association, subchannel assignment, power allocation and phase shifts design were optimized jointly by maximizing the achievable sum rate. 
However, the above works assumed static channels. A downlink SISO RIS-NOMA system over fading channels was considered in~\cite{guo2020intelligent}, where the joint optimization problem over resource allocation and phase shifts was solved by maximizing the average sum rate. 
For the MISO RIS-NOMA systems, to minimize the total transmit power, an alternating difference-of-convex algorithm was proposed by jointly optimizing the active beamforming vectors and phase shift matrix in~\cite{fu2019reconfigurable}.
The total transmit power minimization problem was also studied in~\cite{li2019joint}, where an effective second-order cone programming alternating direction method of multipliers based algorithm was proposed. To further reduce the computational complexity, a low-complexity zero-forcing based algorithm was proposed.
To achieve the capacity region of the MISO RIS-NOMA system, the total transmit power minimization problem was investigated in~\cite{zhu2019power}. An efficient algorithm based on the semidefinite relaxation (SDR) method was proposed to ensure that the considered system achieves the capacity region with high possibility.
In~\cite{mu2019exploiting}, the active beamforming and passive beamforming were optimized jointly by maximizing the system sum rate. The formulated non-convex problem was solved by the alternating optimization, semidefinite programming (SDP) and successive convex approximation approaches.
 \subsection{Motivation and Contributions}
 	Inspired by the aforementioned potential benefits of the RIS and CNOMA, it is interesting to investigate the promising applications of the RIS technique in CNOMA systems for further performance improvement. In light of the above background and to the best of our knowledge, the RIS assisted CNOMA (RIS-CNOMA) system design has not been studied yet. We investigate the potential of the RIS-CNOMA system and identify the key factors affecting the system performance. In particular, we explore the answers to the following questions:
\begin{itemize}
	\item Will the RIS-CNOMA system bring performance gains in term of the total transmit power compared to the conventional CNOMA system without RIS?
   \item With the aid of RIS, which transmission mode is more efficient, cooperative mode (RIS-CNOMA system) or non-cooperative mode (RIS-NOMA system)?
  \item How do the locations of the RIS and users influence the performance of the RIS-CNOMA and RIS-NOMA systems?
\end{itemize}

In this paper, we consider the joint active beamforming vectors, transmit-relaying power and phase shifts optimization problem in the RIS-CNOMA system. 
The main contributions of this paper are summarized as follows:
\begin{enumerate}
	\item We propose a downlink RIS-CNOMA communication system, where the NOMA-stong user acts as a HD mode relay to transmit signal to the NOMA-weak user with the aid of RIS. We formulate the total transmit power minimization problem by jointly optimizing the active beamforming vectors, transmit-relaying power and phase shifts. The formulated problem is a mixed-integer nonlinear programming (MINLP) problem, which is non-trivial to solve directly.
	\item To solve this challenging optimization problem, 
	we first decompose the original problem into two subproblems, i.e, direct transmission (DT)-stage optimization and cooperative transmission (CT)-stage optimization. 
	By utilizing their special structures, we solve the two subproblems optimally. Thus, we obtain the optimal active beamforming vectors, transmit-relaying power and phase shifts.	
	Finally, we propose an alternating optimization based optimal algorithm to update the obtained optimal solutions.  
	\item We develop two low-complexity algorithms to solve the discrete DT-stage and CT-stage phase shifts optimization problems by utilizing the iterative penalty function based SDP and the successive refinement approaches, respectively. Then, we propose a low-complexity alternating optimization based suboptimal algorithm to solve the original MINLP problem.  
	\item Numerical results unveil that 1)  with the aid of RIS, our proposed RIS-CNOMA system outperforms the conventional CNOMA system without the RIS in terms of total transmit power. 2) our proposed suboptimal algorithm is capable of achieving a near-optimal performance. 3) simulation results reveal that the locations of the RIS and users have a significant influence on the performances of the RIS-CNOMA and RIS-NOMA systems in terms of total transmit power.
\end{enumerate}
\subsection{Organization}
The rest of this paper is organized as follows. In Section II, the system model and the
problem formulation for designing the RIS-CNOMA system are presented. In Sections III and IV, we propose the alternating optimization based optimal algorithm and the low-complexity alternating optimization based suboptimal algorithm to solve the original optimization problem, respectively. Numerical results are presented in Section V, which is followed by the conclusions in Section VI.

Notations: $\mathbb{C}^{M \times 1}$ denotes a complex vector of size \emph{M}, diag(\textbf{x}) denotes a diagonal matrix whose diagonal elements are the corresponding elements in vector \textbf{x}. The $m$-th element of vector $\textbf{x}$ is denoted as $\left[ \mathbf{x} \right] _m$ and the $(m,n)$-th element of matrix $\textbf{X}$ is denoted as $\left[ \mathbf{X} \right] _{m,n}$. ${\textbf{x}}^{H}$ denotes the complex conjugate transpose of vector \textbf{x}. The notations Tr(\textbf{X}) and rank(\textbf{X}) denote the trace and rank of matrix \textbf{X}, respectively, while $\angle x$ denotes the phase of a complex number \emph{x}. The functions $\mathcal{R}\left( x \right) $ and $\mathcal{I}\left( x \right) 
$ denote the real and imaginary parts of a complex number \emph{x}. $\mathcal{C}\mathcal{N}\left( 0,\sigma ^2 \right) $ represents a random vector following the distribution of zero mean and $\sigma ^2$ variance .
 \section{System Model and Problem Formulation}
 \subsection{Signal Model}
\begin{figure}[!t]
		 \setlength{\belowcaptionskip}{-0.5cm}   
		\centering
		\includegraphics[scale=0.18]{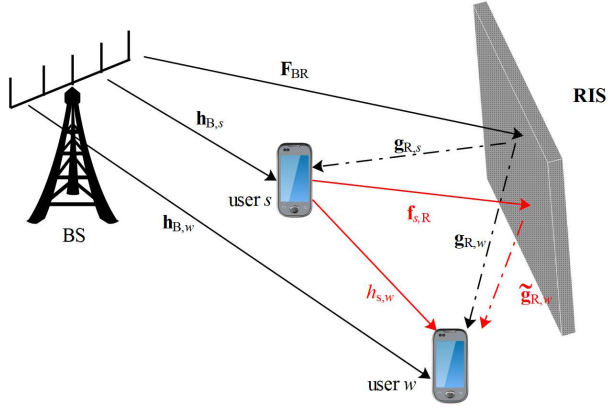}
		\caption{Illustration of the downlink RIS-CNOMA system.}
		\label{system model}
	\end{figure}

	We consider downlink transmissions in an RIS-CNOMA system, as shown in Fig.~\ref{system model}, where the BS is equipped with ${N_{\rm T}}$ antennas, the RIS is equipped with ${L_{\rm RIS}}$ passive reflecting elements, and there are two single-antenna users, denoted by user $s$ and user $w$. Without loss of generality, we assume that user $s$ has a better channel condition than user $w$, i.e, user $s$ is a NOMA-strong user and user $w$ is a NOMA-weak user. With the aid of the RIS, the BS serves the two users using the power-domain NOMA technology, meanwhile the NOMA-strong user $s$ acts as a HD mode relay to transmit signal to the NOMA-weak user $w$. We assume that all the perfect channel state information (CSI) is available at the BS. Indeed, it is challenging to obtain perfect CSI for all channels in RIS assisted communication systems. However, this paper can serve as a theoretical system performance benchmark.
	
	There are two stages involved in the RIS-CNOMA transmission, i.e., the DT and CT stages~\cite{liu2016cooperative}. In the DT stage, the BS transmits the superimposed signal to user $s$ and user $w$ based on the NOMA principle. Meanwhile, the superimposed signal is reflected to the two users by the RIS. User $w$ treats the interference from user $s$ as noise and decodes its own signal. In the CT stage, we assume that	successive interference cancellation (SIC) is successfully implemented at user $s$. Thus, user $s$ first decodes the intended signal for user $w$. Then, the decoded user $w$'s signal is canceled from the received signal of user $s$. Finally, user $w$ receives its decoded signals relayed from user $s$ and reflected from the RIS. As a result, the signal reception diversity at user $w$ is enhanced.
	More specifically, the transmitted superposition coding signal at the BS is $\textbf{x} = {\textbf{w}_{ s}}{x_{ s}} + {\textbf{w}_{w}}{x_{ w}}$, where $x_s$ and $x_w$ are i.i.d information bearing signals for user $s$ and user $w$, respectively, such that $\mathbb{E}\left( \left| x_s \right|^2 \right) =\mathbb{E}\left( \left| x_w \right|^2 \right) =1$, $\textbf{w}_s$ and $\textbf{w}_w$ are the corresponding active beamforming vectors.
	\subsubsection{DT Stage}  \ \
	  
	 Denoting by $\textbf{h}_{\text{B},s}\in \mathbb{C}^{N_{\text{T}}\times 1}$,~$\mathbf{F}_{\text{B},\text{R}}\in \mathbb{C}^{L_{\text{RIS}}\times N_{\text{T}}}
	 $,~${\textbf{g}_{{\rm R},s}} \in \mathbb{C}^{L_{\rm RIS} \times 1}$ the channel coefficients vectors from the BS to user $s$, the BS to the RIS, and the RIS to user $s$, respectively. 
	 The observation at user $s$ is  
	\begin{equation}\label{U1 signal}
		y_s=\left( \mathbf{h}_{\text{B},s}^{H}+\mathbf{g}_{\text{R},s}^{H}\mathbf{\Theta }_{\left( 1 \right)}\mathbf{F}_{\text{B},\text{R}} \right) \mathbf{x}+z_s,
	\end{equation}
	where $z_s\sim \mathcal{CN}\left( 0,\sigma_s ^2 \right) $ is the additive white Gaussian noise (AWGN) with zero mean and variance $\sigma_s ^2$,  $\mathbf{\Theta }_{\left( 1 \right)}=\mathrm{diag}\left\{ e^{j\theta _{1}^{\left( 1 \right)}},e^{j\theta _{2}^{\left( 1 \right)}},\cdots ,e^{j\theta _{_{L_{\rm RIS}}}^{\left( 1 \right)}} \right\} 
	 $ is the DT-stage phase-shift matrix of the RIS and $\theta _{m}^{\left( 1 \right)}$ denotes the DT-stage phase shift of the \emph{m}-th reflecting element,  $m=1,2,\cdots ,L_{\text{RIS}}$. Due to the hardware cost, the phase shifts can only be chosen from a finite set of discrete values. In particular, the set of discrete phase shift values for each reflecting element is given by: $\theta _{m}^{\left( 1 \right)} \in \Omega  \buildrel \Delta \over = \left\{ {0,\frac{{2\pi }}{{{Q}}}, \cdots ,\frac{{2\pi \left( {{Q} - 1} \right)}}{{{Q}}}} \right\}$, where $Q=2^B$ and $B$ is the number of resolution bits of discrete phase shifts.

	According to the NOMA principle, the SIC is performed at user $s$. User $s$ first decodes the signal $x_{w}$ intended to user $w$ and then removes this signal to decode its own signal. Therefore, the achievable data rate for user $s$ to decode the signal of user $w$ can be expressed as~\cite{yuan2019energy}
	\begin{equation}\label{Rate U1-U2}
		R_{s\rightarrow w}=\frac{1}{2}\log _2\left( 1+\frac{\lvert \left( \mathbf{g}_{\text{R},s}^{H}\mathbf{\Theta }_{\left( 1 \right)}\mathbf{F}_{\text{B},\text{R}}+\mathbf{h}_{\text{B},s}^{H} \right) \mathbf{w}_w \rvert^2}{\lvert \left( \mathbf{g}_{\text{R},s}^{H}\mathbf{\Theta }_{\left( 1 \right)}\mathbf{F}_{\text{B},\text{R}}+\mathbf{h}_{\text{B},\text{s}}^{H} \right) \mathbf{w}_s \rvert^2+\sigma _{s}^{2}} \right).  
	\end{equation}
	
	After removing the signal $x_w$ from the received signal, user $s$ can decode its own signal with the achievable data rate, given by
	\begin{equation}\label{Rate U1}
		R_{{s}}=\frac{1}{2}\log _2\left( 1+\frac{\lvert \left( \mathbf{g}_{\text{R},s}^{H}\mathbf{\Theta }_{\left( 1 \right)}\mathbf{F}_{\text{BR}}+\mathbf{h}_{\text{B},s}^{H} \right) \mathbf{w}_s \rvert^2}{\sigma _{s}^{2}} \right).  
	\end{equation}
	
	On the other hand, the received signal at user $w$ in the DT stage is given by
	\begin{equation}\label{signal U2 in phase 1}
		y_{w}^{\left( 1 \right)}=\left( \mathbf{g}_{\text{R},w}^{H}\mathbf{\Theta }_{\left( 1 \right)}\boldsymbol{\rm F}_{\text{B},\text{R}} +\mathbf{h}_{\text{B},w}^{H} \right) \mathbf{x}+z_{w}^{\left( 1 \right)},
	\end{equation}	
where $\mathbf{h}_{\text{B},w}\in \mathbb{C}^{N_{\text{T}}\times 1}$ and $\mathbf{g}_{\text{R},w}\in \mathbb{C}^{L_{\text{RIS}}\times 1}
$ are the channel coefficients vectors from the BS to user $w$ and the RIS to user $w$, respectively, and $z_{w}^{\left( 1 \right)}\sim \mathcal{CN}\left( 0,\sigma _{w}^{2} \right)$ is AWGN with zero mean and variance $\sigma_w ^2$ at the DT stage.
	
The received signal-to-interference-plus-noise-ratio (SINR) at user $w$ to detect its own information can be expressed as
 \begin{equation}\label{SINR_w in phase 1}
	{\rm SINR}_{w}^{\left( 1 \right)}=\frac{\lvert \left( \mathbf{g}_{\text{R},w}^{H}\mathbf{\Theta }_{\left( 1 \right)}\mathbf{F}_{\text{B},\text{R}}+\mathbf{h}_{\text{B},w}^{H} \right) \mathbf{w}_w \rvert^2}{\lvert \left( \mathbf{g}_{\text{R},w}^{H}\mathbf{\Theta }_{\left( 1 \right)}\mathbf{F}_{\text{B},\text{R}}+\mathbf{h}_{\text{B},w}^{H} \right) \mathbf{w}_s \rvert^2+\sigma _{w}^{2}}. 
 \end{equation}	
\subsubsection{CT Stage} \ \

During this stage, user $s$ forwards the decoded signal $x_w$ to user $w$. The observation at user $w$ is given by
	\begin{equation}\label{signal U2 in Phase 2}
		y_{w}^{\left( 2 \right)}=\sqrt{P_{\rm S}}\left(  \widetilde{\mathbf{g}}_{\text{R},w}^{H}\mathbf{\Theta }_{\left( 2 \right)}\boldsymbol{\rm f}_{\text{s},R} + h_{s,w}\right) x_w+z_{w}^{\left( 2 \right)},
	\end{equation}
where $P_{\rm S}$ is the transmit-relaying power at user $s$, $h_{s,w}\in \mathbb{C}^{1\times 1}$, $\boldsymbol{\rm f}_{s,\text{R}}\in \mathbb{C}^{L_{\text{RIS}}\times 1}
$ and $\widetilde{\mathbf{g}}_{{\rm R},w}\in \mathbb{C}^{L_{\text{RIS}}\times 1}$ are the channel coefficients from user $s$ to user $w$, user $s$ to the RIS and the RIS to user $w$, respectively, $z_{w}^{\left( 2 \right)}\sim \mathcal{CN}\left( 0,\sigma _{w}^{2} \right)$ is AWGN with zero mean and variance $\sigma_w ^2$, $\mathbf{\Theta }_{\left( 2 \right)}=\mathrm{diag}\left\{ e^{j\theta _{1}^{\left( 2 \right)}},e^{j\theta _{2}^{\left( 2 \right)}},\cdots ,e^{j\theta _{L_{\text{RIS}}}^{\left( 2 \right)}} \right\}$ is the CT-stage phase-shift matrix of the RIS, and $\theta _{m}^{\left( 2 \right)}\in \Omega$ is the $m$-th CT-stage phase shift, $m=1,2,\cdots ,L_{\text{RIS}}$. 

We obtain the received SINR at user $w$ in the CT stage as follows:
 \begin{equation}\label{SINR_w in P2}
	{\rm SINR}_{w}^{\left( 2 \right)}=\frac{P_{\rm S}\lvert \widetilde{\mathbf{g}}_{\text{R},w}^{H}\mathbf{\Theta }_{\left( 2 \right)}\mathbf{f}_{\text{s},R}+h_{s,w} \rvert^2}{\sigma _{w}^{2}}.
\end{equation}

At the end of the CT stage, user $w$ decodes the signal $x_{w}$ jointly based on the signals received from the BS and user $s$ by using maximal ratio combining (MRC). Therefore, the achievable data rate of user $w$ is given as~\cite{liu2017hybrid,yuan2019energy}
 \begin{equation}\label{Rate U2 MRC}
 	R_{\text{MRC},w}=\frac{1}{2} \log _2\left( 1+\text{SINR}_{w}^{\left( 1 \right)}+\text{SINR}_{w}^{\left( 2 \right)} \right) .
 \end{equation}
	
 Based on the above analysis, the final achievable data rate of user $w$ can be written as~\cite{dinh2020low}
 \begin{equation}\label{Rate U2 final}
 {R_{w}} = \min \left( {R_{s \to w},R_{{\rm MRC},w}} \right).
 \end{equation}
 \subsection{Problem Formulation}
	Our aim is to minimize the total transmit power while guaranteeing the minimum QoS requirements for user $s$ and user $w$. Therefore, we formulate the joint optimization problem over the active beamforming vectors, transmit-relaying power and phase shifts as
\begin{subequations}\label{OOP1}
	 \setlength{\abovedisplayskip}{5pt}
	\begin{align}
	&\mathop {\min }\limits_{{{\textbf{w}}_s},{{\textbf{w}}_w},\boldsymbol{\theta }_{\left( 1 \right)},{P_{\rm S}},\boldsymbol{\theta }_{\left( 2 \right)}} \left\| {{{\textbf{w}}_s}} \right\|_2^2 + \left\| {{{\textbf{w}}_w}} \right\|_2^2 + {P_{\rm S}},   \\
	&{s.t.} \ R_s\ge R_{s}^{\min}, \label{OOP1 b}\\
	&   \ \ \ \ \ R_{s\rightarrow w}\ge R_{w}^{\min},  \label{OOP1 c}\\
	&   \ \ \ \ \ R_{\text{MRC},w}\ge R_{w}^{\min}
	, \label{OOP1 d}\\
	&   \ \ \ \ \ \theta _{m}^{\left( 1 \right)}\in \Omega ,m=1,2,\cdots ,L_{\text{RIS}}, \label{OOP1 e} \\
	&   \ \ \ \ \ \theta _{m}^{\left( 2 \right)}\in \Omega ,m=1,2,\cdots ,L_{\text{RIS}}, \label{OOP1 f}
	\end{align}
\end{subequations}
where $\boldsymbol{\theta }_{\left( 1 \right)}=\left[ \theta _{1}^{\left( 1 \right)},\theta _{2}^{\left( 1 \right)},\cdots ,\theta _{L_{\text{RIS}}}^{\left( 1 \right)} \right] ^T
$ and $\boldsymbol{\theta }_{\left( 2 \right)}=\left[ \theta _{1}^{\left( 2 \right)},\theta _{2}^{\left( 2 \right)},\cdots ,\theta _{L_{\text{RIS}}}^{\left( 2 \right)} \right] ^T
$ are the phase-shift vectors of the RIS in the DT and CT stages, respectively. Constraint~\eqref{OOP1 b} and constraints~\eqref{OOP1 c}-\eqref{OOP1 d} ensure that user $s$ and user $w$ satisfy the minimum QoS requirements $R_{s}^{\min }$ and $R_{w}^{\min }$, respectively.
	
Problem~\eqref{OOP1} is a highly intractable non-convex problem. There are two main challenges to solve problem~\eqref{OOP1}. First, the constraints~\eqref{OOP1 e} and~\eqref{OOP1 f} restrict the DT-stage and CT-stage phase shifts to be discrete values. Therefore, problem~\eqref{OOP1} is a MINLP problem which is generally difficult to solve. Second, the active beamforming vectors, transmit-relaying power and RIS phase shifts are highly coupled, which makes the problem even more challenging. In the following sections, we will develop an alternating optimization based optimal algorithm to decouple the optimization variables and the optimal solutions are achieved in each alternative procedure. In addition, a low-complexity alternating optimization based suboptimal algorithm is also proposed. 
\section{Alternating Optimization based Optimal Algorithm Development}	
In this section, an alternating optimization based optimal algorithm is proposed to solve problem~\eqref{OOP1}, where the active beamforming vectors $\left\{ \mathbf{w}_s,\mathbf{w}_w \right\}$, DT-stage phase shifts $\left\{ \theta _{m}^{\left( 1 \right)} \right\} $, transmit-relaying power $P_{\rm S}$ and CT-stage phase shifts $\left\{ \theta _{m}^{\left( 2 \right)} \right\} $ are optimized alternatively until convergence. To make problem~\eqref{OOP1} tractable, we first decouple it into two subproblems, i.e, DT-stage optimization and CT-stage optimization, which are given by~\eqref{JAP} and~\eqref{JPP}, respectively:
\begin{subequations}\label{JAP}
	  \setlength{\abovedisplayskip}{5pt}
	\begin{align}
	&\mathop {\min} \limits_{\mathbf{w}_s,\mathbf{w}_w,\boldsymbol{\theta }_{\left( 1 \right)}}\lVert \mathbf{w}_s \rVert _{2}^{2}+\lVert \mathbf{w}_w \rVert _{2}^{2},   \\
	&{s.t.} \ ~\eqref{OOP1 b},~\eqref{OOP1 c},~\eqref{OOP1 d},~\eqref{OOP1 e}. \label{BOP C1}
	\end{align}
\end{subequations}	
\begin{subequations}\label{JPP}
	 \setlength{\abovedisplayskip}{-15pt}
	  \setlength{\belowdisplayskip}{5pt}
	\begin{align}
	&\mathop {\min} \limits_{P_{\text{S}},\boldsymbol{\theta }_{\left( 2 \right)}}P_{\text{S}},   \\
	&{s.t.} \ ~\eqref{OOP1 d},~\eqref{OOP1 f}. \label{Power allocation problem b}  
	\end{align}
\end{subequations}	

 Subproblem~\eqref{JAP} is formulated to solve the active beamforing vectors and phase shifts optimization problem in the DT stage with given transmit-relaying power and DT-stage phase shifts. Subproblem~\eqref{JPP} is formulated to solve the transmit-relaying power and CT-stage phase shifts optimization problem in the CT stage with given active beamforming vectors and DT-stage phase shifts. In the following subsections, we will propose efficient algorithms to solve the above two subproblems.
\subsection{Optimization for DT Stage}	
In this subsection, we optimize the active beamforming vectors $\textbf{w}_s$ and $\textbf{w}_w$, and DT-stage phase shifts $\left\{ \theta _{m}^{\left( 1 \right)} \right\}$. 
Since $\textbf{w}_s$, $\textbf{w}_w$ and $\left\{ \theta _{m}^{\left( 1 \right)} \right\}$ are coupled, it is difficult to solve problem~\eqref{JAP} directly. By utilizing the alternating optimization approach, we alternatively optimize the active beamforming vectors and DT-stage phase shifts.
\subsubsection{Optimizing active beamforming vectors $\textbf{w}_s$ and $\textbf{w}_w$} \ \
	
 Define ${\overline {\textbf{h}} _{{{\rm B}},s}} = {\textbf{g}}_{{{\rm R}},s}^H{\bf{\Theta }}{{\textbf{F}}_{{\rm{BR}}}} + {\textbf{h}}_{{\rm{B}},s}^H$, ${\overline {\textbf{h}} _{{\rm{B}},w}} = {\textbf{g}}_{{\rm{R}},w}^H{\bf{\Theta }}{{\textbf{F}}_{{\rm{BR}}}} + {\textbf{h}}_{{\rm{B}},w}^H$ and $\overline{h}_{s,w}=\widetilde{\mathbf{g}}_{\text{R},w}^{H}\mathbf{\Theta }_{\left( 2 \right)}\mathbf{f}_{s,\text{R}}+h_{s,w}
$, then the active beamforming vectors optimization problem with given $\left\{ \theta _{m}^{\left( 1 \right)} \right\} $, $P_{\rm S}$ and $\left\{ \theta _{m}^{\left( 2 \right)} \right\} $ is formulated as
\begin{subequations}\label{ABOP-1}
	 \setlength{\abovedisplayskip}{5pt}
	\begin{align}
	&\mathop {\min }\limits_{{{\textbf{w}}_s},{{\textbf{w}}_w}} \left\| {{{\textbf{w}}_s}} \right\|_2^2 + \left\| {{{\textbf{w}}_w}} \right\|_2^2, \label{ABOP-1 a}  \\
	& {s.t.} \ \lvert \overline{\mathbf{h}}_{{\rm B},s}\mathbf{w}_s \rvert^2\ge r_{s}^{\min}\sigma _{s}^{2}, \label{ABOP-1 b} \\
	&   \ \ \ \ \ \lvert \overline{\mathbf{h}}_{{\rm B},s}\mathbf{w}_w \rvert^2\ge r_{w}^{\min}\left( \lvert \overline{\mathbf{h}}_{{\rm B},s}\mathbf{w}_s \rvert^2+\sigma _{s}^{2} \right) , \label{ABOP-1 c} \\
	&   \ \ \ \ \ \lvert \overline{\mathbf{h}}_{{\rm B},w}\mathbf{w}_w \rvert^2\ge \left( r_{w}^{\min}-\frac{P_{\rm S}\lvert \overline{h}_{s,w} \rvert^2}{\sigma _{w}^{2}} \right) \left( \lvert \overline{\mathbf{h}}_{{\rm B},w}\mathbf{w}_s \rvert^2+\sigma _{w}^{2} \right) , \label{ABOP-1 d}
	\end{align}
\end{subequations}
where $r_k^{\min } = {2^{R_k^{\min }}} - 1$,  $k \in \left\{ {s,w} \right\}$.

By applying the SDR method, we define $\overline{\mathbf{H}}_{\text{B},i}=\overline{\mathbf{h}}_{\text{B},i}^{H}\overline{\mathbf{h}}_{\text{B},i}
$ and ${\textbf {W}_k} = {{\textbf{w}}_k}{\textbf{w}}_k^H$ with ${\rm rank}\left( {{\textbf{W}_k}} \right) = 1$ and $\mathbf{W}_k\ge \textbf{0}$, $ i \in \left\{ {s,w} \right\}$ and $ k \in \left\{ {s,w} \right\}$. We further drop the rank-one constraint due to its non-convexity. Then, the relaxed form of problem~\eqref{ABOP-1} is given by
 	\begin{subequations}\label{ABOP-trace}
	  \setlength{\abovedisplayskip}{5pt}
		\begin{align}
		&\mathop {\min }\limits_{{\textbf{W}_{s}},{\textbf{W}_{w}}} {\rm Tr}\left( {{\textbf{W}_{s}}} \right) + {\rm Tr}\left( {{\textbf{W}_{w}}} \right), \label{ABOP-trace a}  \\
		& {s.t.} \  {\rm Tr}\left( \mathbf{W}_s\overline{\mathbf{H}}_{{\rm B},s} \right) \ge r_{s}^{\min}\sigma _{s}^{2}, \label{ABOP-trace b} \\
		&   \ \ \ \ \ {\rm Tr}\left( \mathbf{W}_w\overline{\mathbf{H}}_{{\rm B},s} \right) \ge r_{w}^{\min}\left( {\rm Tr}\left( \mathbf{W}_s\overline{\mathbf{H}}_{{\rm B},s} \right) +\sigma _{s}^{2} \right) 
		, \label{ABOP-trace c} \\
		&   \ \ \ \ \ {\rm Tr}\left( {{\textbf{W}_w}{{\overline {\textbf H} }_{{\rm{B}},w}}} \right) \ge \left( {r_w^{\min } - \frac{{{P_s}{{\left| {{{\overline h }_{s,w}}} \right|}^2}}}{{\sigma _w^2}}} \right)\left( {{\rm Tr}\left( {{\textbf{W}_s}{{\overline {\textbf H} }_{{\rm{B}},w}}} \right) + \sigma _w^2} \right), \label{ABOP-trace d} \\
		&   \ \ \ \ \ \mathbf{W}_s\ge \textbf{0}, \mathbf{W}_w\ge \textbf{0}. \label{ABOP-trace e}
		\end{align}
	\end{subequations}
 
 It is noted that problem~\eqref{ABOP-trace} is a standard SDP problem and can be solved efficiently by using the SDP solver in CVX~\cite{cvx}. If $\mathbf{W}_{w}$ and $\mathbf{W}_{s}$ satisfy ${\rm rank}\left( {{\textbf{W}_s}} \right) = 1$ and ${\rm rank}\left( {{\textbf{W}_w}} \right) = 1$, then the optimal active beamforming vectors $\mathbf{w}_{s}$ and $\mathbf{w}_{s}$ of problem~\eqref{ABOP-1} can be obtained from the eigenvalue decomposition (EVD) of $\textbf{W}_s$ and $\textbf{W}_w$, respectively~\cite{xu2017joint}. Otherwise, if there exists beamforming matrices such that ${\rm rank}\left( {{\textbf{W}_s}} \right) > 1$ and ${\rm rank}\left( {{\textbf{W}_w}} \right) > 1$, then the solutions $\mathbf{W}_{s}$ and $\mathbf{W}_{s}$ are not optimal for problem~\eqref{ABOP-1} in general. The following proposition shows that the optimal solution of problem~\eqref{ABOP-trace} satisfies ${\rm rank}\left( {{\textbf{W}_s}} \right) = 1$ and ${\rm rank}\left( {{\textbf{W}_w}} \right) = 1$.
 
\begin{proposition}\label{P-rank1}
	The optimal solution to problem~\eqref{ABOP-trace} without the rank-one constraint always satisfies ${\rm rank}\left( {{\textbf{W}_s}} \right) = 1$ and ${\rm rank}\left( {{\textbf{W}_w}} \right) = 1$.
\end{proposition} 
	
\textit{Proof}: According to Theorem 3.2 in~\cite{huang2010rank-constrained}, we conclude that the SDR problem~\eqref{ABOP-trace} always has optimal solution $\mathbf{W}_{s}$ and $\mathbf{W}_{w}$, which satisfy the relation 
\begin{equation}\label{rank+rank}
  \setlength{\abovedisplayskip}{5pt}
	{\rm rank}{^2}\left( {{\textbf{W}_s}} \right) + {\rm rank}{^2}\left( {{\textbf{W}_w}} \right) \le 3.
\end{equation}

To meet the QoS constraints~\eqref{ABOP-trace b},~\eqref{ABOP-trace c} and~\eqref{ABOP-trace d}, the optimal $\textbf{W}_s \ne \textbf{0}$ and $\textbf{W}_w \ne \textbf{0}$. Therefore, there must exist an optimal solution with ${\rm rank}\left( {{\textbf{W}_s}} \right) = 1$ and ${\rm rank}\left( {{\textbf{W}_w}} \right) = 1$.
\begin{remark}
	\textbf{Proposition~\ref{P-rank1}} shows that the rank relaxation  is always tight and the optimal solution of the SDR problem~\eqref{ABOP-trace} is the same as the original problem~\eqref{ABOP-1}. 
\end{remark}

 \subsubsection{Optimizing DT-stage phase shifts $\left\{ \theta _{m}^{\left( 1 \right)} \right\}$ } \ \
 
 With given  $\textbf{w}_s$, $\textbf{w}_w$, $P_{\rm S}$ and $\left\{ \theta _{m}^{\left( 2 \right)} \right\} $, the DT-stage phase shifts optimization problem is formulated as 
\begin{subequations}\label{Phase1OP-rewritten}
	 \setlength{\abovedisplayskip}{5pt}
	\begin{align}
	&{\rm find}~\boldsymbol{\theta }_{\left( 1 \right)}, \label{PBOP-rewritten a} \\
	&s.t.   \lvert \left( \mathbf{g}_{\text{R},s}^{H}\mathbf{\Theta }_{\left( 1 \right)}\mathbf{F}_{\text{B},\text{R}}+\mathbf{h}_{\text{B},s}^{H} \right) \mathbf{w}_s \rvert^2\ge r_{s}^{\min}\sigma _{s}^{2}, \label{PBOP-rewritten1 b} \\
	&   \ \ \ \  \lvert \left( \mathbf{g}_{\text{R},s}^{H}\mathbf{\Theta }_{\left( 1 \right)}\mathbf{F}_{\text{B},\text{R}}+\mathbf{h}_{\text{B},s}^{H} \right) \mathbf{w}_w \rvert^2\ge r_{w}^{\min}\left( \lvert \left( \mathbf{g}_{\text{R},s}^{H}\mathbf{\Theta F}_{\text{B},\text{R}}+\mathbf{h}_{\text{B},s}^{H} \right) \mathbf{w}_s \rvert^2+\sigma _{s}^{2} \right), \label{PBOP-rewritten1 c} \\
	&   \ \ \ \  \lvert \left( \mathbf{g}_{\text{R},w}^{H}\mathbf{\Theta }_{\left( 1 \right)}\mathbf{F}_{\text{B},\text{R}}+\mathbf{h}_{\text{B},w}^{H} \right) \mathbf{w}_w \rvert^2\ge \eta \left( \lvert \left( \mathbf{g}_{\text{R},w}^{H}\mathbf{\Theta }_{\left( 1 \right)}\mathbf{F}_{\text{B},\text{R}}+\mathbf{h}_{\text{B},w}^{H} \right) \mathbf{w}_s \rvert^2+\sigma _{w}^{2} \right), \label{PBOP-rewritten1 d} \\
	&   \ \ \ \ \eqref{OOP1 e}, \label{PBOP-rewritten1 e}
	\end{align}
\end{subequations}
where $\eta =r_{w}^{\min}-\frac{P_{\rm S}\lvert \widetilde{\mathbf{g}}_{\text{R},w}^{H}\mathbf{\Theta }_{\left( 2 \right)}\mathbf{f}_{s,\text{R}}+h_{s,w} \rvert^2}{\sigma _{w}^{2}}$.

Problem~\eqref{Phase1OP-rewritten} is a MINLP problem, which is difficult to solve directly. To make problem~\eqref{Phase1OP-rewritten} tractable, we reformulate it as an ILP problem. Denoting by $\mathbf{v}_{\left( 1 \right)}=\left[ e^{j\theta _{1}^{\left( 1 \right)}}e^{j\theta _{2}^{\left( 1 \right)}}\cdots e^{j\theta _{L_{\rm RIS}}^{\left( 1 \right)}} \right] ^H$ the passive beamforming vector, then we have
 \begin{equation}\label{combined channel1 rewritten}
	\lvert \left( \mathbf{g}_{\text{R},i}^{H}\mathbf{\Theta F}_{\text{B},\text{R}}+\mathbf{h}_{\text{B},i}^{H} \right) \mathbf{w}_k \rvert^2=\mathbf{v}_{\left( 1 \right)}^{H}\mathbf{X}_{i,k}\mathbf{v}_{\left( 1 \right)}+2\mathcal{R}\left( \mathbf{v}_{\left( 1 \right)}^{H}y_{i,k} \right) +\mathbf{h}_{\text{B},i}^{H}\mathbf{W}_k\mathbf{h}_{\text{B},i},
 \end{equation}
 where ${{\textbf X}_{i,k}} = {\rm diag}\left( {{\textbf{g}}_{{\rm{R,}}i}^H} \right){{\textbf{F}}_{{\rm{B,R}}}}{\textbf{W}_k}{\textbf{F}}_{{\rm{B,R}}}^H{\rm diag}\left( {{{\textbf{g}}_{{\rm{R,}}i}}} \right)$,
 ${{\textbf{y}}_{i,k}} = {\rm diag}\left( {{\textbf{g}}_{{\rm{R,}}i}^H} \right){{\textbf{F}}_{{\rm{B,R}}}}{\textbf{W}_k}{{\textbf{h}}_{{\rm{B,}}i}}$, $k \in \left\{ {s,w} \right\}$, $i \in \left\{ {s,w} \right\}$.
 
 With the above variable definitions, problem~\eqref{Phase1OP-rewritten} can be rewritten as 
	\begin{subequations}\label{PBOP-rewritten2}
	  \setlength{\abovedisplayskip}{5pt}
	\begin{align}
	&{\rm find}~\mathbf{v}_{\left( 1 \right)}, \label{PBOP-rewritten a} \\
	&s.t.  \mathbf{v}_{\left( 1 \right)}^{H}\mathbf{X}_{s,s}\mathbf{v}_{\left( 1 \right)}+2\mathcal{R}\left( \mathbf{v}_{\left( 1 \right)}^{H}\mathbf{y}_{s,s} \right) \ge r_{s}^{\min}\sigma _{s}^{2}-\mathbf{h}_{\text{B},s}^{H}\mathbf{W}_s\mathbf{h}_{\text{B},s}, \label{PBOP-rewritten1 b} \\
	&   \ \ \ \ \mathbf{v}_{\left( 1 \right)}^{H}\left( \mathbf{X}_{s,w}-r_{w}^{\min}\mathbf{X}_{s,s} \right) \mathbf{v}_{\left( 1 \right)}+2\mathcal{R}\left( \mathbf{v}_{\left( 1 \right)}^{H}\left( \mathbf{y}_{s,w}-r_{w}^{\min}\mathbf{y}_{s,s} \right) \right) \ge \alpha _1
	 , \label{PBOP-rewritten1 c} \\
	&   \ \ \ \ \mathbf{v}_{\left( 1 \right)}^{H}\left( \mathbf{X}_{w,w}-\eta \mathbf{X}_{w,s} \right) \mathbf{v}_{\left( 1 \right)}+2\mathcal{R}\left( \mathbf{v}_{\left( 1 \right)}^{H}\left( \mathbf{y}_{w,w}-\eta \mathbf{y}_{w,s} \right) \right) \ge \alpha _2, \label{PBOP-rewritten1 d} \\
	&   \ \ \ \ \eqref{OOP1 e}, \label{PBOP-rewritten1 e}
	\end{align}
\end{subequations}	 
 where $\alpha _1=\mathbf{h}_{\text{B},s}^{H}\left( r_{w}^{\min}\mathbf{W}_s-\mathbf{W}_w \right) \mathbf{h}_{\text{B},s}+r_{w}^{\min}\sigma _{s}^{2}
 $ and $\alpha _2=\mathbf{h}_{\text{B},w}^{H}\left( \eta \mathbf{W}_s-\mathbf{W}_w \right) \mathbf{h}_{\text{B},w}+\eta \sigma _{w}^{2}
 $.
 
 It is noted that problem~\eqref{PBOP-rewritten2} is still non-convex. However, by exploiting its special structure, problem~\eqref{PBOP-rewritten2} can be further reformulated as an ILP problem~\cite{wu2020beamforming}, whose globally optimal solution can be obtained. We introduce the following proposition to transform problem~\eqref{PBOP-rewritten2} to an ILP problem with only binary optimization variables. 
\begin{proposition}\label{Integer expression}
 Define the new auxiliary vector ${{\boldsymbol \theta} _\Delta } = {\left[ { - {{\frac{{\left( {Q - 1} \right)2\pi }}{Q}}_{}}{ \cdots _{}} - {{\frac{{2\pi }}{Q}}_{}}{0_{}}{{\frac{{2\pi }}{Q}}_{}}{ \cdots _{}}\frac{{\left( {Q - 1} \right)2\pi }}{Q}} \right]^T}$, and introduce two binary indicator vectors $\boldsymbol{\delta }_{m}^{\left( 1 \right)}
 $ and $\widetilde{\boldsymbol{\delta }}_{m,n}^{\left( 1 \right)} $, whose q-th elements, denoted by $\left[ \boldsymbol{\delta }_{m}^{\left( 1 \right)} \right] _q
 $ and $\left[ \widetilde{\boldsymbol{\delta }}_{m,n}^{\left( 1 \right)} \right] _q$, indicate whether phase shift $\theta _{m}^{\left( 1 \right)}=\left[ \boldsymbol{\theta }_{\Delta} \right] _q$ and phase-shift difference $\theta _{m}^{\left( 1 \right)}-\theta _{n}^{\left( 1 \right)}=\left[ \boldsymbol{\theta }_{\Delta} \right] _q$, respectively. Then, we have: $\theta _{m}^{\left( 1 \right)}=\boldsymbol{\theta }_{\Delta}^{T}\boldsymbol{\delta }_{m}^{\left( 1 \right)}
 $ and $\theta _{m}^{\left( 1 \right)}-\theta _{n}^{\left( 1 \right)}=\boldsymbol{\theta }_{\Delta}^{T}\widetilde{\boldsymbol{\delta }}_{m,n}^{\left( 1 \right)}$. Thus, problem~\eqref{PBOP-rewritten2} can be re-expressed as
 \begin{subequations}\label{phase 1 MIL problem}
 	 \setlength{\abovedisplayskip}{5pt}
 	\begin{align}
 	&{\rm find} \left\lbrace {\boldsymbol \delta}_{\left( 1 \right) }, ~\widetilde {\boldsymbol \delta}_{\left( 1 \right) } \right\rbrace , \label{MIL problem a} \\
 	&s.t.\psi \left( \mathbf{X}_{s,s},\mathbf{y}_{s,s},\boldsymbol{\delta }_{\left( 1 \right)},\widetilde{\boldsymbol{\delta }}_{\left( 1 \right)} \right) \ge r_{s}^{\min}\sigma _{s}^{2}-\mathbf{h}_{\text{B},s}^{H}\mathbf{W}_s\mathbf{h}_{\text{B},s}, \label{MIL problem b} \\
 	&   \ \ \ \ \psi \left( \mathbf{X}_{s,w}-r_{w}^{\min}\mathbf{X}_{s,s},\mathbf{y}_{s,w}-r_{w}^{\min}\mathbf{y}_{s,s},\boldsymbol{\delta }_{\left( 1 \right)},\widetilde{\boldsymbol{\delta }}_{\left( 1 \right)} \right) \ge \mathbf{h}_{\text{B},s}^{H}\left( r_{w}^{\min}\mathbf{W}_s-\mathbf{W}_w \right) \mathbf{h}_{\text{B},s}+r_{w}^{\min}\sigma _{s}^{2}	, \label{MIL problem c} \\
 	&   \ \ \ \ \psi \left( \mathbf{X}_{w,w}-\eta \mathbf{X}_{w,s},\mathbf{y}_{w,w}-\eta \mathbf{y}_{w,s},\boldsymbol{\delta }_{\left( 1 \right)},\widetilde{\boldsymbol{\delta }}_{\left( 1 \right)} \right) \ge \mathbf{h}_{\text{B},w}^{H}\left( \eta \mathbf{W}_s-\mathbf{W}_w \right) \mathbf{h}_{\text{B},w}+\eta \sigma _{w}^{2}, \label{MIL problem d} \\
 	&   \ \ \ \   \left[ \boldsymbol{\delta }_{m}^{\left( 1 \right)} \right] _q=\left\{ 0,1 \right\} ,\sum_{q=Q}^{2Q-1}{\left[ \boldsymbol{\delta }_{m}^{\left( 1 \right)} \right] _q}=1,\sum_{q=1}^{Q-1}{\left[ \boldsymbol{\delta }_{m}^{\left( 1 \right)} \right] _q}=0,
 	\label{MIL problem e}\\
 	&   \ \ \ \  \left[ \widetilde{\boldsymbol{\delta }}_{m,n}^{\left( 1 \right)} \right] _q\in \left\{ 0,1 \right\} ,\sum_{q=0}^{2Q-1}{\left[ \widetilde{\boldsymbol{\delta }}_{m,n}^{\left( 1 \right)} \right] _q}=1,\label{MIL problem f}\\
 	&   \ \ \ \ \boldsymbol{\theta }_{\Delta}^{T}\boldsymbol{\delta }_{m}^{\left( 1 \right)}-\boldsymbol{\theta }_{\Delta}^{T}\boldsymbol{\delta }_{n}^{\left( 1 \right)}=\boldsymbol{\theta }_{\Delta}^{T}\widetilde{\boldsymbol{\delta }}_{m,n}^{\left( 1 \right)},\label{MIL problem g}
 	\end{align}
 \end{subequations}	 
 with 
\begin{equation}\label{Linear express}
\begin{array}{l}
\psi \left( \mathbf{A},\mathbf{b},\boldsymbol{\delta }_{\left( 1 \right)},\widetilde{\boldsymbol{\delta }}_{\left( 1 \right)} \right) 
\\
=2\sum\limits_{m=1}^{L_{\rm{RIS}}-1}{\sum\limits_{n=m+1}^{L_{\rm{RIS}}}{\mathcal{R}\left\{ \left[ \mathbf{A} \right] _{m,n}\left( \mathbf{v}_{sum}^{T}\widetilde{\boldsymbol{\delta }}_{m,n}^{\left( 1 \right)} \right) \right\}}}+2\mathcal{R}\left( \left[ \mathbf{b} \right] _m\mathbf{v}_{sum}^{T}\boldsymbol{\delta }_{m}^{\left( 1 \right)} \right) +\sum\limits_{m=1}^{L_{\rm{RIS}}}{\left[ \mathbf{A} \right] _{m,m}}
\end{array},
\end{equation} 
 where  $\boldsymbol{\delta }_{\left( 1 \right)}=\left\{ \boldsymbol{\delta }_{m}^{\left( 1 \right)},m=1,2,\cdots ,L_{\text{RIS}} \right\} $, $\widetilde{\boldsymbol{\delta }}_{\left( 1 \right)}=\left\{ \widetilde{\boldsymbol{\delta }}_{m}^{\left( 1 \right)},m=1,2,\cdots ,L_{\text{RIS}} \right\} 
 $, $\mathbf{v}_{\rm sum}^{T}=\mathbf{v}_{\cos}^{T}+j\mathbf{v}_{\sin}^{T}
 $, $\mathbf{v}_{\cos}=\left[ \cos \left( \left[ \boldsymbol{\theta }_{\Delta} \right] _0 \right) _{}\cdots _{}\cos \left( \left[ \boldsymbol{\theta }_{\Delta} \right] _{2Q-1} \right) \right] ^T$, $\mathbf{v}_{\sin}=\left[ \sin \left( \left[ \boldsymbol{\theta }_{\Delta} \right] _0 \right) _{}\cdots _{}\sin \left( \left[ \boldsymbol{\theta }_{\Delta} \right] _{2Q-1} \right) \right] ^T$, $\textbf{A} \in {\mathbb{C}^{{L_{\rm RIS}} \times {L_{\rm RIS}}}}$ is a Hermit matrix and $\textbf{b} \in {\mathbb{C}^{{L_{\rm RIS}} \times 1}}$ is column vector.
 \end{proposition}
 
\textit{Proof}: See Appendix A.

It is noted that problem~\eqref{phase 1 MIL problem} is an ILP problem~\cite{wu2020beamforming}, which can be optimally solved using CVX solver, such as the Moseck solver~\cite{cvx}. After solving problem~\eqref{phase 1 MIL problem}, the optimal DT-stage phase shifts $\left\{ \theta _{m}^{\left( 1 \right)} \right\} $ can be obtained through $\theta _{m}^{\left( 1 \right)}=\boldsymbol{\theta }_{\Delta}^{T}\boldsymbol{\delta }_{m}^{\left( 1 \right)}
$, $m = 1,2, \cdots ,L_{\rm RIS}$.
\subsection{Optimization for CT Stage}
In this subsection, we optimize the transmit-relaying power $P_{\rm S}$ and CT-stage RIS phase shifts $\left\{ \theta _{m}^{\left( 2 \right)} \right\}$. According to problem~\eqref{JPP}, the joint transmit-relaying power and CT-stage RIS phase shifts optimization problem can be rewritten as 
\begin{subequations}\label{JPP1}
	\begin{align}
	&\mathop {\min }\limits_{{P_{\rm S}}, \left\{ \theta _{m}^{\left( 2 \right)} \right\} } {P_{\rm S}}, \label{POP a}  \\
	& {s.t.} \ P_{\rm S}\lvert \widetilde{\mathbf{g}}_{\text{R},w}^{H}\mathbf{\Theta }_{\left( 2 \right)}\mathbf{f}_{s,\text{R}}+h_{s,w} \rvert^2\ge \sigma _{w}^{2}\left( r_{w}^{\min}-{\rm SINR}_{w}^{\left( 1 \right)} \right). \label{POP b}
	\end{align}
\end{subequations}

In the following, we will optimize $P_{\rm S}$ and $\left\{ \theta _{m}^{\left( 2 \right)} \right\}$ based on the principle of the alternating optimization. 
\subsubsection{Optimizing transmit-relaying power $P_{\rm S}$} \ \

With given $\textbf{w}_s$, $\textbf{w}_w$, $\boldsymbol{\theta }_{\left( 1 \right)}$ and $\boldsymbol{\theta }_{\left( 2 \right)}$, the transmit-relaying power optimization problem can be reduced to the following problem
\begin{subequations}\label{POP1}
	\begin{align}
	&\mathop {\min }\limits_{{P_{\rm S}}} {P_{\rm S}}, \label{POP a}  \\
	& {s.t.} \ P_s\ge \frac{\sigma _{w}^{2}\left( r_{w}^{\min}-{\rm SINR}_{w}^{\left( 1 \right)} \right)}{\lvert \widetilde{\mathbf{g}}_{\text{R},w}^{H}\mathbf{\Theta }_{\left( 2 \right)}\mathbf{f}_{s,\text{R}}+h_{s,w} \rvert^2}. \label{POP1 d}
	\end{align}
\end{subequations}

Clearly, the optimal solution of problem~\eqref{POP1} is
\begin{equation}\label{optimal power solution}
P_{\text{S}}=\max \left( 0,\frac{\sigma _{w}^{2}\left( r_{w}^{\min}-{\rm SINR}_{w}^{\left( 1 \right)} \right)}{\lvert \widetilde{\mathbf{g}}_{\text{R},w}^{H}\mathbf{\Theta }_{\left( 2 \right)}\mathbf{f}_{s,\text{R}}+h_{s,w} \rvert^2} \right). 
\end{equation}

In the following remark, we provide more insights on the optimal transmit-relaying power in ~\eqref{optimal power solution}.
\begin{remark}\label{optimal power reamrk}
	According to Equation~\eqref{optimal power solution}, if ${\rm SINR}_{w}^{\left( 1 \right)}\geqslant r_{w}^{\min}
	$,  then $P_{\rm S}=0$. If ${\rm SINR}_{w}^{\left( 1 \right)}<r_{w}^{\min}$, then $P_{\text{S}}=\frac{\sigma _{w}^{2}\left( r_{w}^{\min}-{\rm SINR}_{w}^{\left( 1 \right)} \right)}{\lvert \widetilde{\mathbf{g}}_{\text{R},w}^{H}\mathbf{\Theta }_{\left( 2 \right)}\mathbf{f}_{s,\text{R}}+h_{s,w} \rvert^2}$. The first case implies that when
	the achievable SINR of user $w$ satisfies the minimum QoS requirement in the DT stage, then there is no need for user $s$ to cooperate with user $w$. For the second case, when the achievable SINR of user $w$ in the DT stage is less than the minimum QoS requirement, user $s$ needs to cooperate with user $w$ to improve the performance of user $w$. Therefore, the obtained optimal transmit-relaying power~\eqref{optimal power solution} for problem~\eqref{POP1} is in agreement with our intuition.
\end{remark}
 \subsubsection{Optimizing CT-Stage phase shifts $\left\{ \theta _{m}^{\left( 2 \right)} \right\}$ } \ \
 
It is worth pointing out that when $P_S=0$, there is no need to optimize the CT-stage RIS phase shifts. In the following, we will only consider the CT-stage RIS phase shifts optimization under $P_{\text{S}}>0 $. With given $\textbf{w}_s$, $\textbf{w}_w$, $\boldsymbol{\theta }_{\left( 1 \right)}$ and $P_{\rm S}$, the CT-stage RIS phase shifts optimization problem is formulated as 	
\begin{subequations}\label{P2OP-rewritten}
	\begin{align}
	&{\rm find}~\boldsymbol{\theta }_{\left( 2 \right)}, \label{PBOP2-rewritten a} \\
	&s.t.  \  \lvert \widetilde{\mathbf{g}}_{\text{R},w}^{H}\mathbf{\Theta }_{\left( 2 \right)}\mathbf{f}_{s,\text{R}}+h_{s,w} \rvert^2\ge \frac{\sigma _{w}^{2}{\rho }}{P_s}, \label{PBOP2-rewritten b} \\
	&   \ \ \ \ \ \theta _{m}^{\left( 2 \right)}\in \Omega ,m=1,2,\cdots ,L_{\text{RIS}}, \label{PBOP2-rewritten c}
	\end{align}
\end{subequations}
where ${\rho }={r}_{w}^{\min}-\frac{\lvert \left( \mathbf{g}_{\text{R},w}^{H}\mathbf{\Theta }_{\left( 1 \right)}\mathbf{F}_{\text{B},\text{R}}+\mathbf{h}_{\text{B},w}^{H} \right) \mathbf{w}_w \rvert^2}{\lvert \left( \mathbf{g}_{\text{R},w}^{H}\mathbf{\Theta }_{\left( 1 \right)}\mathbf{F}_{\text{B},\text{R}}+\mathbf{h}_{\text{B},w}^{H} \right) \mathbf{w}_s \rvert^2+\sigma _{w}^{2}}
$.

 Problem~\eqref{P2OP-rewritten} is also an MINLP problem. Similar to solving problem~\eqref{PBOP-rewritten2}, problem~\eqref{P2OP-rewritten} can also be transformed to an ILP problem. Define $\mathbf{Z}_w=\mathrm{diag}\left( \widetilde{\mathbf{g}}_{\text{R},w}^{H} \right) \mathbf{f}_{s,\text{R}}\mathbf{f}_{s,\text{R}}^{H}\mathrm{diag}\left( \widetilde{\mathbf{g}}_{\text{R},w} \right) 
 $ and $\mathbf{u}_w=\mathrm{diag}\left( \widetilde{\mathbf{g}}_{\text{R},w}^{H} \right) \mathbf{f}_{s,\text{R}}h_{s,w}^{H}$, then solving problem~\eqref{P2OP-rewritten} is equivalent to solving the following problem 
 \begin{subequations}\label{PBOP2-rewritten2}
 	\begin{align}
 	&{\rm find}~\boldsymbol{\theta }_{\left( 2 \right)}, \label{PBOP2-rewritten a} \\
 	&s.t.~\mathbf{v}_{\left( 2 \right)}^{H}\mathbf{Z}_w\mathbf{v}_{\left( 2 \right)}+2\mathcal{R}\left( \mathbf{v}_{\left( 2 \right)}^{H}\mathbf{u}_w \right) \ge \frac{\sigma _{w}^{2}{\rho }}{P_s}-\lvert h_{s,w} \rvert^2, \label{PBOP2-rewritten b} \\
 	&   \ \ \ \ \ \theta _{m}^{\left( 2 \right)}\in \Omega ,m=1,2,\cdots ,L_{\text{RIS}}. \label{PBOP2-rewritten c}
 	\end{align}
 \end{subequations}

We introduce two binary indicator vectors  $\boldsymbol{\delta }_{m}^{\left( 2 \right)}
$ and $\widetilde{\boldsymbol{\delta }}_{m,n}^{\left( 2 \right)} $, where $\left[ \boldsymbol{\delta }_{m}^{\left( 2 \right)} \right] _q$ and $\left[ \widetilde{\boldsymbol{\delta }}_{m,n}^{\left( 2 \right)} \right] _q$ indicate whether $\theta _{m}^{\left( 2 \right)}=\left[ \boldsymbol{\theta }_{\Delta} \right] _q$ and $\theta _{m}^{\left( 2 \right)}-\theta _{n}^{\left( 2 \right)}=\left[ \boldsymbol{\theta }_{\Delta} \right] _q$, respectively. Based on the above definitions, we have:  $\theta _{m}^{\left( 2 \right)}=\boldsymbol{\theta }_{\Delta}^{T}\boldsymbol{\delta }_{m}^{\left( 2 \right)}
$ and $\theta _{m}^{\left( 2 \right)}-\theta _{n}^{\left( 2 \right)}=\boldsymbol{\theta }_{\Delta}^{T}\widetilde{\boldsymbol{\delta }}_{m,n}^{\left( 1 \right)}$. Furthermore, the introduced binary indicator vectors $\boldsymbol{\delta }_{m}^{\left( 2 \right)}
$ and $\widetilde{\boldsymbol{\delta }}_{m,n}^{\left( 2 \right)} $ should satisfy the following constraints:
\begin{equation}\label{binary 2 constraint 1}
\left[ \boldsymbol{\delta }_{m}^{\left( 2 \right)} \right] _q=\left\{ 0,1 \right\} ,\sum_{q=Q}^{2Q-1}{\left[ \boldsymbol{\delta }_{m}^{\left( 2 \right)} \right] _q}=1,\sum_{q=1}^{Q-1}{\left[ \boldsymbol{\delta }_{m}^{\left( 2 \right)} \right] _q}=0,
\end{equation}
\begin{equation}\label{binary 2 constraint 2}  
\left[ \widetilde{\boldsymbol{\delta }}_{m,n}^{\left( 2 \right)} \right] _q\in \left\{ 0,1 \right\} ,\sum_{q=0}^{2Q-1}{\left[ \widetilde{\boldsymbol{\delta }}_{m,n}^{\left( 2 \right)} \right] _q}=1,
\end{equation} 
\begin{equation}\label{binary 2 constraint 3} 
\boldsymbol{\theta }_{\Delta}^{T}\boldsymbol{\delta }_{m}^{\left( 2 \right)}-\boldsymbol{\theta }_{\Delta}^{T}\boldsymbol{\delta }_{n}^{\left( 2 \right)}=\boldsymbol{\theta }_{\Delta}^{T}\widetilde{\boldsymbol{\delta }}_{m,n}^{\left( 2 \right)},
\end{equation} 
where $m,n = 1,2, \cdots ,{L_{\rm RIS}}$ and $q = 0,1, \cdots ,\left( {2Q - 1} \right)$.

 According to \textbf{Proposition~\ref{Integer expression}}, problem~\eqref{PBOP2-rewritten2} can be reformulated as an ILP problem, given by
\begin{subequations}\label{phase 2 MIL problem}
	\begin{align}
	&{\rm find} \left\{ \boldsymbol{\delta }_{\left( 2 \right)},\widetilde{\boldsymbol{\delta }}_{\left( 2 \right)} \right\} 
	, \label{MIL problem2 a} \\
	&s.t.  \ \psi \left( \mathbf{Z}_w,\mathbf{u}_w,\boldsymbol{\delta }_{\left( 2 \right)},\widetilde{\boldsymbol{\delta }}_{\left( 2 \right)} \right) \ge \frac{\sigma _{w}^{2}\rho}{P_s}-\lvert h_{s,w} \rvert^2
	, \label{MIL problem2 b} \\
	&   \ \ \ \ ~\eqref{binary 2 constraint 1},~\eqref{binary 2 constraint 2}, ~\eqref{binary 2 constraint 3}, 
	\end{align}
\end{subequations}	
where $\boldsymbol{\delta }_{\left( 2 \right)}=\left\{ \boldsymbol{\delta }_{m}^{\left( 2 \right)},m=1,2,\cdots ,L_{\text{RIS}} \right\} 
$ and $\widetilde{\boldsymbol{\delta }}_{\left( 2 \right)}=\left\{ \widetilde{\boldsymbol{\delta }}_{m}^{\left( 2 \right)},m=1,2,\cdots ,L_{\text{RIS}} \right\} 
$.
\subsection{Proposed Algorithm, Convergence and Complexity}
To facilitate the understanding of the proposed alternating optimization based optimal (AOBO) algorithm, we summarize it in
\textbf{Algorithm~\ref{Optimal Algorithm}}.
It should be noted that if the obtained transmit-relaying power  $P_{\rm S} =0$, user $s$ will not relay signal to user $w$ and there is also no signal reflected by the RIS. Thus, CT-stage RIS phase shifts will not be optimized. Otherwise, if $P_{\rm S}>0$, we need to optimize CT-stage RIS phase shifts.
\begin{algorithm}
	\caption{Alternating Optimization Based Optimal (AOBO) Algorithm}
	\label{Optimal Algorithm}
	\begin{algorithmic}[1]
		\STATE  Initialize $\boldsymbol{\theta }_{\left( 1 \right)}^{\left( 0 \right)}$, ${{{P}}_{\rm S}^{\left( 0 \right)}}$ and $\boldsymbol{\theta }_{\left( 2 \right)}^{\left( 0 \right)}$. Set the iteration index ${t} = 1$.
		\REPEAT
		\STATE  update ${{\textbf{w}}_s^{\left( t \right)}} $ and ${{\textbf{w}}_w^{\left( t \right)}} $ by solving problem~\eqref{ABOP-trace} with $\boldsymbol{\theta }_{\left( 1 \right)}^{\left( t-1 \right)}$, ${{{P}}_{\rm S}^{\left( t-1 \right)}}$ and $\boldsymbol{\theta }_{\left( 2 \right)}^{\left( t-1 \right)}$;
	   \STATE  update $\boldsymbol{\theta }_{\left( 1 \right)}^{\left( t \right)}$ by solving problem~\eqref{phase 1 MIL problem} with $ {{\textbf{w}}_s^{\left( t \right)}}$, $ {{\textbf{w}}_w^{\left( t \right)}}$, ${{{P}}_{\rm S}^{\left( {t-1} \right)}}$ and $\boldsymbol{\theta }_{\left( 2 \right)}^{\left( t-1 \right)}$;
		\STATE  update ${{{P}}_{\rm S}^{\left( {t} \right)}}$ according to formulation~\eqref{optimal power solution} with  $ {{\textbf{w}}_s^{\left( t \right)}}$, $ {{\textbf{w}}_w^{\left( t \right)}}$, $\boldsymbol{\theta }_{\left( 1 \right)}^{\left( t \right)}$ and $\boldsymbol{\theta }_{\left( 2 \right)}^{\left( t-1 \right)}$;
		\STATE  \textbf{if}  ${{{P}}_{\rm S}^{\left( {t} \right)}}=0$
		\STATE      \ \ \ \   go to step 11;
		 \STATE  \textbf{else}
		 \STATE   \ \ \ \    update $\boldsymbol{\theta }_{\left( 2 \right)}^{\left( t \right)}$ by solving problem~\eqref{phase 2 MIL problem} with  $ {{\textbf{w}}_s^{\left( t \right)}}$, $ {{\textbf{w}}_w^{\left( t \right)}}$, $\boldsymbol{\theta }_{\left( 1 \right)}^{\left( t \right)}$, and  ${{{P}}_{\rm S}^{\left( {t} \right)}}$;   
	  \STATE  \textbf{end if}
		\STATE  ${t} = {t} + 1$;
		\UNTIL {the objective value of problem~\eqref{OOP1} converge.}
		\STATE   \textbf{Output}: optimal$ {{\textbf{w}}_s^{\left( t \right)}}$, $ {{\textbf{w}}_w^{\left( t \right)}}$, $\boldsymbol{\theta }_{\left( 1 \right)}^{\left( t \right)}$, ${{{P}}_{\rm S}^{\left( {t} \right)}}$ and $\boldsymbol{\theta }_{\left( 2 \right)}^{\left( t \right)}$.
	\end{algorithmic}
\end{algorithm}
 
\subsubsection{Convergence analysis}\ \

Let $\mathcal{F}\left( \left\{ \mathbf{w}_{k}^{\left( {t} \right)} \right\} ,\boldsymbol{\theta }_{\left( 1 \right)}^{\left( t \right)},P_{{\rm S}}^{\left( t \right)},\boldsymbol{\theta }_{\left( 2 \right)}^{\left( t \right)} \right) $ denote the objective value of problem~\eqref{OOP1} in the $t$-th iteration, where $\left\{ \mathbf{w}_{k}^{\left( t \right)} \right\}$, $ \boldsymbol{\theta }_{\left( 1 \right)}^{\left( t \right)}$, $P_{{\rm S}}^{\left( t \right)}$ and $\boldsymbol{\theta }_{\left( 2 \right)}^{\left( t \right)}
$ are the $t$-th iteration solutions, $k\in \left\{ w,s \right\} 
$. To begin with, we have:
\begin{equation}\label{converge equation} 
\begin{split}
	\mathcal{F}\left( \left\{ \mathbf{w}_{k}^{\left( {t}-1 \right)} \right\} ,\boldsymbol{\theta }_{\left( 1 \right)}^{\left( t-1 \right)},P_{\text{S}}^{\left( t-1 \right)},\boldsymbol{\theta }_{\left( 2 \right)}^{\left( t-1 \right)} \right) 
 & \overset{\left( a \right)}{\geqslant}\mathcal{F}\left( \left\{ \mathbf{w}_{k}^{\left( {t} \right)} \right\} ,\boldsymbol{\theta }_{\left( 1 \right)}^{\left( t-1 \right)},P_{\text{S}}^{\left( t-1 \right)},\boldsymbol{\theta }_{\left( 2 \right)}^{\left( t-1 \right)} \right) 
   \\
 & \overset{\left( b \right)}{=}\mathcal{F}\left( \left\{ \mathbf{w}_{k}^{\left( {t} \right)} \right\} ,\boldsymbol{\theta }_{\left( 1 \right)}^{\left( t \right)},P_{\text{S}}^{\left( t-1 \right)},\boldsymbol{\theta }_{\left( 2 \right)}^{\left( t-1 \right)} \right) 
   \\
& \overset{\left( c \right)}{\geqslant}\mathcal{F}\left( \left\{ \mathbf{w}_{k}^{\left( {t} \right)} \right\} ,\boldsymbol{\theta }_{\left( 1 \right)}^{\left( t \right)},P_{\text{S}}^{\left( t \right)},\boldsymbol{\theta }_{\left( 2 \right)}^{\left( t-1 \right)} \right) 
   \\
&\overset{\left( d \right)}{=}\mathcal{F}\left( \left\{ \mathbf{w}_{k}^{\left( {t} \right)} \right\} ,\boldsymbol{\theta }_{\left( 1 \right)}^{\left( t \right)},P_{\text{S}}^{\left( t \right)},\boldsymbol{\theta }_{\left( 2 \right)}^{\left( t \right)} \right) 
\end{split},
\end{equation} 
where (a) holds since for given $\boldsymbol{\theta }_{\left( 1 \right)}^{\left( t-1 \right)}$, $P_{\text{S}}^{\left( t-1 \right)}$ and $\boldsymbol{\theta }_{\left( 2 \right)}^{\left( t-1 \right)}$, $\left\{ \mathbf{w}_{k}^{\left( t \right)} \right\}$ are the optimal solutions to problem~\eqref{OOP1};
 (b) holds because the objective function of problem~\eqref{OOP1} has nothing to do with $\boldsymbol{\theta }_{\left( 1 \right)}$, and only depends on $\left\{ \textbf{w}_k \right\} $ and $P_{\rm S}$; (c) comes from the fact that $P_{\rm S}^{\left( t \right)}$ is the optimal solution to problem~\eqref{OOP1} with given $\left\{ \mathbf{w}_{k}^{\left( \boldsymbol{t} \right)} \right\}$, $\boldsymbol{\theta }_{\left( 1 \right)}^{\left( t \right)}$  and $\boldsymbol{\theta }_{\left( 2 \right)}^{\left( t-1 \right)}$; (d) is obtained since the objective function of problem~\eqref{OOP1} has nothing to do with $\boldsymbol{\theta }_{\left( 2 \right)}$, and only depends on $\left\{ \textbf{w}_k \right\} $ and $P_{\rm S}$;

The inequality in~\eqref{converge equation} indicates that the objective value of problem~\eqref{OOP1} is monotonically non-increasing after each iteration. On the other hand, the total transmit power is lower bounded due to the minimum QoS requirement for each user. Therefore, the proposed algorithm is guaranteed to converge.
\subsubsection{Complexity analysis} \ \

The complexity of \textbf{Algorithm~\ref{Optimal Algorithm}} mainly depends on the complexities for solving problem~\eqref{ABOP-trace}, problem~\eqref{phase 1 MIL problem} and problem~\eqref{phase 2 MIL problem}. In particular, the complexity for solving the SDP problem~\eqref{ABOP-trace} is $O\left( 2N_{\rm T}^{6} \right)$. In general, the worst case computational complexity for solving an ILP problem is exponential in the number of variables, which can be approximated as $O\left( 2^N \right) $, where $N$ is the number of binary variables. It is not difficult to see that the number of binary variables in problem~\eqref{phase 1 MIL problem} is $2Q\left( L_{\rm RIS}^{2}+L_{\rm RIS} \right) $. Therefore, the complexity of solving problem~\eqref{phase 1 MIL problem} is $O\left( 2^{2Q\left( L_{\rm RIS}^{2}+L_{\rm RIS} \right)} \right) $. Similarly, the complexity of solving the ILP problem~\eqref{phase 2 MIL problem} is also $O\left( 2^{2Q\left( L_{\rm RIS}^{2}+L_{\rm RIS} \right)} \right) $. Thus, the total complexity of \textbf{Algorithm~\ref{Optimal Algorithm}} is $O\left( t_{\max}\left( 2N_{\rm T}^{6}+2^{2Q\left( L_{\rm RIS}^{2}+L_{\rm RIS} \right) +1} \right) \right) $, where $t_{\max}$ denotes the iteration number.
\section{Low-Complexity Alternating Optimization Based Suboptimal Algorithms} 

In this section, we propose two low-complexity suboptimal algorithms to solve the DT-stage and CT-stage RIS phase shifts optimization problems, respectively . 
 \vspace{-0.3cm}
\subsection{Low-Complexity SubOptimal Algorithm for DT-Stage RIS Phase Shifts Optimization} 
To transform problem~\eqref{Phase1OP-rewritten} to a more tractable form, define ${\textbf{E}_{{\rm B},i}} = \left[ {\begin{array}{*{20}{l}}
	{{\rm diag}\left( {{\textbf{g}}_{{\rm{R,}}i}^H} \right){{\textbf{F}}_{{\rm{B,R}}}}}\\
	{{\textbf{h}}_{{\rm{B}},i}^H}
	\end{array}} \right]$, $\mathbf{q}_{s,w}=\left[ \begin{array}{c}
	\mathrm{diag}\left( \widetilde{\mathbf{g}}_{\text{R},w}^{H} \right) \mathbf{f}_{s,R}\\
	h_{s,w}\\
\end{array} \right]$  and $\overline{\mathbf{v}}_{\left( 1 \right)}=\left[ e^{j\theta _{1}^{\left( 1 \right)}}e^{j\theta _{2}^{\left( 1 \right)}}\cdots e^{j\theta _{L_{\rm RIS}}^{\left( 1 \right)}}1 \right] ^H$. By substituting the above variable definitions into problem~\eqref{Phase1OP-rewritten}, then we have
\begin{subequations}\label{Phase1OP-r2}
	\begin{align}
	&{\rm find}~~\overline{\mathbf{v}}_{\left( 1 \right)}, \label{Phase1OP-r2 a} \\
	&s.t.\lvert \overline{\mathbf{v}}_{\left( 1 \right)}^{H}\mathbf{E}_{{\rm B},s}\mathbf{w}_s \rvert^2\ge r_{s}^{\min}\sigma _{s}^{2}, \label{Phase1OP-r2 b} \\
	&   \ \ \ \ \lvert \overline{\mathbf{v}}_{\left( 1 \right)}^{H}\mathbf{E}_{{\rm B},s}\mathbf{w}_w \rvert^2\ge r_{w}^{\min}\left( \lvert \overline{\mathbf{v}}_{\left( 1 \right)}^{H}\mathbf{E}_{{\rm B},s}\mathbf{w}_s \rvert^2+\sigma _{s}^{2} \right) , \label{Phase1OP-r2 c} \\
	&   \ \ \ \ \lvert \overline{\mathbf{v}}_{\left( 1 \right)}^{H}\mathbf{E}_{{\rm B},w}\mathbf{w}_w \rvert^2\ge \eta \left( \lvert \overline{\mathbf{v}}_{\left( 1 \right)}^{H}\mathbf{E}_{{\rm B},w}\mathbf{w}_s \rvert^2+\sigma _{w}^{2} \right), \label{Phase1OP-r2 d}  \\
	&   \ \ \ \ \left| \left[ \overline{\mathbf{v}}_{\theta} \right] _m \right|=1, m=1,2,\cdots ,L_{{\rm RIS}}+1, \label{Phase1OP-r2 e} 
	\end{align}
\end{subequations}
where $\eta =r_{w}^{\min}-P_{\rm S}{\rm Tr}\left( \overline{\textbf{v}}_{\left( 2 \right)}^{H}\mathbf{q}_{s,w} \right) $ and $\overline{\mathbf{v}}_{\left( 2 \right)}=\left[ e^{j\theta _{1}^{\left( 2 \right)}}e^{j\theta _{2}^{\left( 2 \right)}}\cdots e^{j\theta _{L_{\rm RIS}}^{\left( 2 \right)}}1 \right] ^H$.

Define $\overline{\mathbf{V}}_{\left( 1 \right)}=\overline{\mathbf{v}}_{\left( 1 \right)}\overline{\mathbf{v}}_{\left( 1 \right)}^{H}
$, which satisfies $\overline{\mathbf{V}}_{\left( 1 \right)}\geqslant 0$ and ${\rm rank}\left( \overline{\mathbf{V}}_{\left( 1 \right)} \right) = 1$, then problem~\eqref{Phase1OP-r2} can be rewritten as
\begin{subequations}\label{POP trace-rank2}
	\begin{align}
	&{\rm find}~\overline{\mathbf{V}}_{\left( 1 \right)}, \label{POP trace-rank1 a} \\
	&s.t. \ {\rm Tr}\left( \mathbf{E}_{{\rm B},s}\mathbf{W}_s\mathbf{E}_{{{\rm B},s}}^{H}\overline{\mathbf{V}}_{\left( 1 \right)} \right) \ge r_{s}^{\min}\sigma _{s}^{2}, \label{POP trace-rank1 b} \\
	&   \ \ \ \ \  {\rm Tr}\left( \mathbf{E}_{{\rm B},s}\mathbf{W}_w\mathbf{E}_{{{\rm B},s}}^{H}\overline{\mathbf{V}}_{\left( 1 \right)} \right) \ge r_{w}^{\min}\left( {\rm Tr}\left( \mathbf{E}_{{\rm B},s}\mathbf{W}_s\mathbf{E}_{{{\rm B},s}}^{H}\overline{\mathbf{V}}_{\left( 1 \right)} \right) +\sigma _{s}^{2} \right) 
	, \label{POP trace-rank1 c} \\
	&   \ \ \ \ \ {\rm Tr}\left( \mathbf{E}_{{\rm B},w}\mathbf{W}_w\mathbf{E}_{{{\rm B},w}}^{H}\overline{\mathbf{V}}_{\left( 1 \right)} \right) \ge \eta \left( {\rm Tr}\left( \mathbf{E}_{{\rm B},w}\mathbf{W}_s\mathbf{E}_{{{\rm B},w}}^{H}\overline{\mathbf{V}}_{\left( 1 \right)} \right) +\sigma _{w}^{2} \right) 
	, \label{POP trace-rank1 d} \\
	&   \ \ \ \ \ \left[  \overline{\mathbf{V}}_{\left( 1 \right)} \right]_{m,m} =1,m=1,2,\cdots ,L_{\text{RIS}}+1, \label{POP trace-rank1 e} \\
	&   \ \ \ \ \	\overline{\mathbf{V}}_{\left( 1 \right)}\geqslant \textbf{0}, \label{POP trace-rank1 f}\\
	&   \ \ \ \ \ {\rm rank}\left( {{\overline{\mathbf{V}}_{\left( 1 \right)}}} \right) = 1. \label{POP trace-rank1 g}
	\end{align}
\end{subequations}	 

However, problem~\eqref{POP trace-rank2} is still a non-convex problem due to the non-covex rank-one constraint~\eqref{POP trace-rank1 g}. To tackle the rank-one constraint, we first rewrite the constraint~\eqref{POP trace-rank1 g} as~\cite{lin2019joint,lin2019robust}
\begin{equation}\label{equivalent rank constraint}
{\rm Tr}\left( \overline{\mathbf{V}}_{\left( 1 \right)} \right) -\varepsilon _{\max}\left( \overline{\mathbf{V}}_{\left( 1 \right)} \right) \le 0,
\end{equation} 
where $\varepsilon _{\max}\left( \overline{\mathbf{V}}_{\left( 1 \right)} \right)$ denotes the maximum eigenvalue of matrix ${{{\overline {\textbf V} }_{\left( 1 \right)} }}$.

The rank-one constraint ${\rm rank}\left( {{\overline{\mathbf{V}}_{\left( 1 \right)}}} \right) = 1$ implies that ${\rm Tr}\left( \overline{\mathbf{V}}_{\left( 1 \right)} \right) =\varepsilon _{\max}\left( \overline{\mathbf{V}}_{\left( 1 \right)} \right) $ and $\overline{\mathbf{V}}_{\left( 1 \right)}$ only has one non-zero eigenvalue. Thus, we have:
$\overline{\mathbf{V}}_{\left( 1 \right)}=\varepsilon _{\max}\left( \overline{\mathbf{V}}_{\left( 1 \right)} \right) \overline{\mathbf{v}}_{\max}\overline{\mathbf{v}}_{\max}^{H}$, where $\overline{\mathbf{v}}_{\max}$ is the eigenvector corresponding to the maximum eigenvalue. If we make the value of ${\rm Tr}\left( \overline{\mathbf{V}}_{\left( 1 \right)} \right) -\varepsilon _{\max}\left( \overline{\mathbf{V}}_{\left( 1 \right)} \right) $ as small as possible, $\overline{\mathbf{V}}_{\left( 1 \right)}
$ can be approximated by $\varepsilon _{\max}\left( \overline{\mathbf{V}}_{\left( 1 \right)} \right) \overline{\mathbf{v}}_{\max}\overline{\mathbf{v}}_{\max}^{H}$. Based on the above analysis, let  ${\rm Tr}\left( \overline{\mathbf{V}}_{\left( 1 \right)} \right) -\varepsilon _{\max}\left( \overline{\mathbf{V}}_{\left( 1 \right)} \right) $ be a penalty function and substitute it into the objective function of problem~\eqref{POP trace-rank2}, then we have
\begin{subequations}\label{POP trace}
	\begin{align}
	&\mathop {\min }\limits_{{{\overline {\textbf V} }_{\left( 1 \right)} }} {\rm Tr}\left( \overline{\mathbf{V}}_{\left( 1 \right)} \right) -\varepsilon _{\max}\left( \overline{\mathbf{V}}_{\left( 1 \right)} \right) , \label{POP trace a} \\
	&s.t. ~\eqref{POP trace-rank1 b},~\eqref{POP trace-rank1 c},~\eqref{POP trace-rank1 d},~\eqref{POP trace-rank1 e},~\eqref{POP trace-rank1 f}.
	\end{align}
\end{subequations}	 

It is noted that ${\varepsilon _{\max }}\left( {{{\overline {\textbf V} }_{\left( 1 \right)} }} \right)$ is not differentiable, i.e, non-smooth. Since the subgradient of $ \varepsilon _{\max}\left( \overline{\mathbf{V}}_{\left( 1 \right)} \right)$ is $\overline{\mathbf{v}}_{\max}\overline{\mathbf{v}}_{\max}^{H} $, by utilizing the first order approximation, $\varepsilon _{\max}\left( \overline{\mathbf{V}}_{\left( 1 \right)} \right) $ can be approximated as~\cite{phan2012nonsmooth}
\begin{equation}\label{maxeigen inequality}
\varepsilon _{\max}\left( \overline{\mathbf{V}}_{\left( 1 \right)} \right) \geqslant \varepsilon _{\max}\left( \overline{\mathbf{V}}_{\left( 1 \right)}^{\left( \tau_1 \right)} \right) +{\rm Tr}\left( \overline{\mathbf{v}}_{\max}^{\left( \tau_1 \right)}\left( \overline{\mathbf{v}}_{\max}^{\left( \tau_1 \right)} \right) ^H\left( \overline{\mathbf{V}}_{\left( 1 \right)}-\overline{\mathbf{V}}_{\left( 1 \right)}^{\left( \tau_1 \right)} \right) \right), 
\end{equation} 
where $\overline{\mathbf{V}}_{\left( 1 \right)}^{\left( \tau_1 \right)}$ and $\overline{\mathbf{v}}_{\max}^{\left( \tau_1 \right)} $ are the values of $\overline{\mathbf{V}}_{\left( 1 \right)} $ and $\overline{\mathbf{v}}_{\max}
$ at the $\tau_1$-th iteration, respectively.

Substituting~\eqref{maxeigen inequality} into the objective function of problem~\eqref{POP trace} and ignoring all the irrelevant constant terms, problem~\eqref{POP trace} can be rewritten as
\begin{subequations}\label{POP trace3}
	\begin{align}
	&\underset{\overline{\mathbf{V}}_{\left( 1 \right)}}{\min}\left\{ {\rm Tr}\left( \overline{\mathbf{V}}_{\left( 1 \right)} \right) -{\rm Tr}\left( \overline{\mathbf{v}}_{\max}^{\left( \tau_1 \right)}\left( \overline{\mathbf{v}}_{\max}^{\left( \tau_1 \right)} \right) ^H\overline{\mathbf{V}}_{\left( 1 \right)} \right) \right\} , \label{POP trace2 a} \\
	&s.t. ~\eqref{POP trace-rank1 b},~\eqref{POP trace-rank1 c},~\eqref{POP trace-rank1 d},~\eqref{POP trace-rank1 e},~\eqref{POP trace-rank1 f}
	\end{align}
\end{subequations}

Now, it is easy to verify that problem~\eqref{POP trace3} is a standard SDP problem, which can be  solved efficiently using CVX solvers, such as the SDP solver~\cite{cvx}.
After solving problem~\eqref{POP trace3}, we can obtain $\overline{\mathbf{v}}_{\left( 1 \right)}$ by applying the singular value decomposition (SVD) to $\overline{\mathbf{V}}_{\left( 1 \right)}^{\left( \tau_1 \right)}$. Then, the discrete phase shifts $\left\{ \theta _{m}^{\left( 1 \right)} \right\} 
$ can be calculated by
\begin{align}\label{discrete theta solution}
\theta _{m}^{\left( 1 \right)}=\text{arg}\mathop {\min} \limits_{\theta \in \Omega}\lvert \theta -\angle \left( \left[ \overline{\mathbf{v}}_{\left( 1 \right)} \right] _m \right) \rvert.
\end{align}

\textbf{Algorithm~\ref{Low Phase Shifts1 Algorithm}} summarizes the proposed low-compelxity DT-stage phase shifts optimization algorithm to solve problem~\eqref{Phase1OP-rewritten}.
\begin{algorithm}
	\caption{Low-Compelxity DT-Stage Phase Shifts Optimization Algorithm}
	\label{Low Phase Shifts1 Algorithm}
	\begin{algorithmic}[1]
		\STATE  Initialize feasible point $\overline{\mathbf{V}}_{\left( 1 \right)}^{\left( 0 \right)}$ and set the iteration index ${\tau_1} = 1$.
		\REPEAT
		\STATE  update $\overline{\mathbf{V}}_{\left( 1 \right)}^{\left( \tau_1 \right)}$ by solving problem~\eqref{POP trace3};
		\STATE  ${\tau_1} = {\tau_1} + 1$;
		\UNTIL {the objective value of problem~\eqref{POP trace} converge.}
		\STATE Calculate$\overline{\mathbf{v}}_{\left( 1 \right)}^{\left( \tau_1 \right)}$ by using SVD to $\overline{\mathbf{V}}_{\left( 1 \right)}^{\left( \tau_1 \right)}$.
		\STATE Calculate phase shifts $\theta _{m}^{\left( 1 \right)}$ via~\eqref{discrete theta solution}, $m=1,2,\cdots 
		,L_{\rm RIS}$.
		\STATE   \textbf{Output}: optimal DT-stage phase shifts $\theta _{m}^{\left( 1 \right)}, m=1,2,\cdots 
		,L_{\rm RIS}$.
	\end{algorithmic}
\end{algorithm}
\subsection{Low-Complexity SubOptimal Algorithm for CT-Stage RIS Phase Shifts Optimization} 
According to Equation~\eqref{optimal power solution}, the optimal transmit-relaying power $P_{\rm S}$ can be rewritten as
\begin{equation}\label{optimal power solution 2}
P_{\text{S}}=\frac{\sigma _{w}^{2}}{\lvert \widetilde{\mathbf{g}}_{\text{R},w}^{H}\mathbf{\Theta }_{\left( 2 \right)}\mathbf{f}_{s,\text{R}}+h_{s,w} \rvert^2}\left( r_{w}^{\min}-\frac{\text{Tr}\left( \mathbf{W}_w\overline{\mathbf{H}}_{\text{B},w} \right)}{\text{Tr}\left( \mathbf{W}_s\overline{\mathbf{H}}_{\text{B},w} \right) +\sigma _{w}^{2}} \right). 
\end{equation}	

It is not difficult to see from Equation~\eqref{optimal power solution 2} that minimizing the transmit-relaying power $P_{\rm S}$ is equivalent to maximizing the combined channel gain $\lvert \widetilde{\mathbf{g}}_{\text{R},w}^{H}\mathbf{\Theta }_{\left( 2 \right)}\mathbf{f}_{s,\text{R}}+h_{s,w} \rvert^2$ with given $\textbf{w}_s$, $\textbf{w}_w$ and $\boldsymbol{\theta }_{\left( 1 \right)}$. Therefore, the CT-stage phase shifts optimization problem~\eqref{P2OP-rewritten} can be reformulated as	
\begin{subequations}\label{PS2 low}
	\begin{align}
	&\underset{\left\{ \theta _{m}^{\left( 2 \right)} \right\}}{\max}~\lvert \widetilde{\mathbf{g}}_{\text{R},w}^{H}\mathbf{\Theta }_{\left( 2 \right)}\mathbf{f}_{s,\text{R}}+h_{s,w} \rvert^2, \label{POP trace2 a} \\
	&s.t. \ \theta _{m}^{\left( 2 \right)}\in \Omega ,m=1,2,\cdots ,L_{\text{RIS}}.
	\end{align}
\end{subequations}

Problem~\eqref{PS2 low} can be further transformed to a more tractable form as  
\begin{subequations}\label{PS2 low rewritten}
	\begin{align}
	&\underset{\left\{ \theta _{m}^{\left( 2 \right)} \right\}}
	{\max}~\mathbf{v}_{\left( 2 \right)}^{H}\mathbf{Z}_w\mathbf{v}_{\left( 2 \right)}+2\mathcal{R}\left( \mathbf{v}_{\left( 2 \right)}^{H}\mathbf{u}_w \right) +\lvert h_{s,w} \rvert^2, \\
	&s.t. \ \theta _{m}^{\left( 2 \right)}\in \Omega ,m=1,2,\cdots ,L_{\text{RIS}}.
	\end{align}
\end{subequations}

Before solving problem~\eqref{PS2 low rewritten}, we first rewrite the objective function of problem~\eqref{PS2 low rewritten} as  
\begin{equation}\label{linear phase n} 
\begin{split}
& \mathbf{v}_{\left( 2 \right)}^{H}\mathbf{Z}_w\mathbf{v}_{\left( 2 \right)}+2\mathcal{R}\left( \mathbf{v}_{\left( 2 \right)}^{H}\mathbf{u}_w \right) +\left| h_{s,w} \right|^2
  \\
 &=2\mathcal{R}\left( \left| \alpha _l \right|e^{j\left( \theta _{l}^{\left( 2 \right)}-\angle \alpha _l \right)} \right) +\sum_{m\ne l}^{L_{\text{RIS}}}{\sum_{n\ne l}^{L_{\text{RIS}}}{\mathbf{Z}_w\left( m,n \right) e^{j\left( \theta _{m}^{\left( 2 \right)}-\theta _{n}^{\left( 2 \right)} \right)}}}+2\mathcal{R}\left( \sum_{m\ne l}^{L_{\text{RIS}}}{e^{j\theta _{m}^{\left( 2 \right)}}\mathbf{u}_w\left( m \right)} \right) 
  \\
 &+\left| h_{s,w} \right|^2+\mathbf{Z}_w\left( l,l \right) 
 \end{split},
 \end{equation} 
where $\alpha _l$ is defined as
\begin{equation}\label{alpha _l}
  \alpha _l=\sum_{n\ne l}^{L_{\text{RIS}}}{\mathbf{Z}_w\left( l,n \right) e^{-j\theta _{n}^{\left( 2 \right)}}}+\mathbf{u}_w\left( l \right) =\left| \alpha _l \right|e^{-j\angle \alpha _l}.
 \end{equation} 

It is noted that the rewritten objective function~\eqref{linear phase n} is linear with respect to $\theta _{l}^{\left( 2 \right)}
$ with fixed $\theta _{m}^{\left( 2 \right)},m\ne l$. Therefore, the successive refinement algorithm~\cite{wu2020beamforming} can be used to solve problem~\eqref{PS2 low rewritten}. By dropping all the irrelevant constant terms, the $l$-th phase shift optimization problem given all the other phase shifts $\theta _{m}^{\left( 2 \right)},m\ne l$ can be formulated as
\begin{subequations}\label{PS3 low}
	\begin{align}
	&~\underset{\theta _{l}^{\left( 2 \right)}}{\max}~\mathcal{R}\left( \left| \alpha _l \right|e^{j\left( \theta _{l}^{\left( 2 \right)}-\angle \alpha _l \right)} \right), 
	 \\
	&s.t. \ \theta _{l}^{\left( 2 \right)}\in \Omega.
	\end{align}
\end{subequations}

It is easy to show that the optimal solution to the above problem is 
\begin{equation}\label{optimal phase 2}
\theta _{l}^{\left( 2 \right)}=arg\underset{\theta \in \Omega}{\min}\left| \theta -\angle \alpha _l \right|. 
\end{equation} 

We can successively obtain all the phase shifts according to~\eqref{optimal phase 2} in the order from $l=1$ to $l=L_{\rm RIS}$ and repeat this procedure until converge. The proposed low-compelxity CT-stage RIS phase shifts optimization algorithm to solve problem~\eqref{PS2 low} is summarized in \textbf{Algorithm~\ref{Low Phase Shifts2 Algorithm}}.
\begin{algorithm}
	\caption{Low-Compelxity CT-Stage RIS Phase Shifts Optimization Algorithm}
	\label{Low Phase Shifts2 Algorithm}
	\begin{algorithmic}[1]
		\STATE  Initialize $\theta _{l}^{\left( 2,0 \right)},l=2,3,\cdots ,L_{\rm RIS}$ and set the iteration index ${\tau_2} = 1$. 
		\REPEAT
		\STATE  Successively update $\theta _{l}^{\left( 2,\tau _2 \right)}$ according to~\eqref{optimal phase 2} in the order from $l=1$ to $l=L_{\rm RIS}$.
		\STATE  ${\tau_2} = {\tau_2} + 1$;
		\UNTIL {the objective value of problem~\eqref{PS2 low rewritten} converge.}   
	\end{algorithmic}
\end{algorithm}
 \vspace{-0.3cm}
\subsection{Proposed Algorithm, Convergence and Complexity}
Based on the above discussions, we provide the details of the proposed low complexity alternating optimization based suboptimal (LCAOBS) algorithm to solve the original problem~\eqref{OOP1}  in \textbf{Algorithm~\ref{overall low comeplexity Algorithm}}.
\begin{algorithm}
	\caption{Low Complexity Alternating Optimization Based Suboptimal (LCAOBS) Algorithm}
	\label{overall low comeplexity Algorithm}
	\begin{algorithmic}[1]
		\STATE  Initialize $\boldsymbol{\theta }_{\left( 1 \right)}^{\left( 0 \right)}$, ${{{P}}_{\rm S}^{\left( 0 \right)}}$ and $\boldsymbol{\theta }_{\left( 2 \right)}^{\left( 0 \right)}$. Set the iteration index ${t} = 1$.
		\REPEAT
		\STATE  update ${{\textbf{w}}_s^{\left( t \right)}} $ and ${{\textbf{w}}_w^{\left( t \right)}} $ by solving problem~\eqref{ABOP-trace} with $\boldsymbol{\theta }_{\left( 1 \right)}^{\left( t-1 \right)}$, ${{{P}}_{\rm S}^{\left( t-1 \right)}}$ and $\boldsymbol{\theta }_{\left( 2 \right)}^{\left( t-1 \right)}$;
		\STATE  update $\boldsymbol{\theta }_{\left( 1 \right)}^{\left( t \right)}$ via \textbf{Algorithm~\ref{Low Phase Shifts1 Algorithm}} with $ {{\textbf{w}}_s^{\left( t \right)}}$, $ {{\textbf{w}}_w^{\left( t \right)}}$, ${{{P}}_{\rm S}^{\left( {t-1} \right)}}$ and $\boldsymbol{\theta }_{\left( 2 \right)}^{\left( t-1 \right)}$;
		\STATE  update ${{{P}}_{\rm S}^{\left( {t} \right)}}$ according to formulation~\eqref{optimal power solution} with $ {{\textbf{w}}_s^{\left( t \right)}}$, $ {{\textbf{w}}_w^{\left( t \right)}}$, $\boldsymbol{\theta }_{\left( 1 \right)}^{\left( t \right)}$ and $\boldsymbol{\theta }_{\left( 2 \right)}^{\left( t-1 \right)}$;
		\STATE  \textbf{if} ${{{P}}_{\rm S}^{\left( {t} \right)}} = 0$
		\STATE      \ \ \ \   go to step 11;
		\STATE  \textbf{else}
		\STATE   \ \ \ \    update $\boldsymbol{\theta }_{\left( 2 \right)}^{\left( t \right)}$ via \textbf{Algorithm~\ref{Low Phase Shifts2 Algorithm}} with  $ {{\textbf{w}}_s^{\left( t \right)}}$, $ {{\textbf{w}}_w^{\left( t \right)}}$, $\boldsymbol{\theta }_{\left( 1 \right)}^{\left( t \right)}$, and ${{{P}}_{\rm S}^{\left( {t} \right)}}$;   
		\STATE  \textbf{end if}
		\STATE  ${t} = {t} + 1$;
		\UNTIL {the objective value of problem~\eqref{OOP1} converge.}
		\STATE   \textbf{Output}: optimal $ {{\textbf{w}}_s^{\left( t \right)}}$, $ {{\textbf{w}}_w^{\left( t \right)}}$, $\boldsymbol{\theta }_{\left( 1 \right)}^{\left( t \right)}$, ${{{P}}_{\rm S}^{\left( {t} \right)}}$ and $\boldsymbol{\theta }_{\left( 2 \right)}^{\left( t \right)}$.
	\end{algorithmic}
\end{algorithm}
\subsubsection{Convergence analysis}
There are two iterative layers in \textbf{Algorithm~\ref{overall low comeplexity Algorithm}}, namely, the inner-layer and outer-layer iterations. \textbf{Algorithms~\ref{Low Phase Shifts1 Algorithm}} and \textbf{\ref{Low Phase Shifts2 Algorithm}} are the inner-layer iterative algorithms. The convergence analyses of \textbf{Algorithms~\ref{Low Phase Shifts1 Algorithm}} and \textbf{\ref{Low Phase Shifts2 Algorithm}} are presented in \textbf{Propositions~\ref{Phase1 algorithm converge}} and \textbf{\ref{Phase2 algorithm converge}}, respectively.         
The convergence analysis of the outer-layer iteration in \textbf{Algorithm~\ref{overall low comeplexity Algorithm}} is similar to the convergence analysis in \textbf{Algorithm~\ref{Optimal Algorithm}}. Since both the inner-layer and the outer-layer iterations converge, the proposed \textbf{Algorithm~\ref{overall low comeplexity Algorithm}} converges.
\begin{proposition}\label{Phase1 algorithm converge} 
The objective value of problem~\eqref{POP trace} is non-increasing over iterations until convergence by applying the proposed iterative
penalty function based SDP algorithm, i.e, \textbf{Algorithm~\ref{Low Phase Shifts1 Algorithm}}.
\end{proposition}	
\textit{Proof}: See Appendix B.
 \begin{proposition}\label{Phase2 algorithm converge} 
 	The objective value of problem~\eqref{PS2 low rewritten} is non-decreasing over iterations until convergence by applying the proposed successive refinement algorithm, i.e, \textbf{Algorithm~\ref{Low Phase Shifts2 Algorithm}}.
 \end{proposition}
\textit{Proof}: See Appendix C.
\subsubsection{Complexity analysis}
In \textbf{Algorithm~\ref{overall low comeplexity Algorithm}}, the complexity of \textbf{Algorithm~\ref{Low Phase Shifts1 Algorithm}} for solving the SDP problem~\eqref{POP trace3} is $O\left( \tau _{1}^{\max}\left( L_{\rm RIS}+1 \right) ^6 \right) $ and the complexity of \textbf{Algorithm~\ref{Low Phase Shifts2 Algorithm}} by using the successive refinement algorithm to solve problem~\eqref{PS2 low rewritten} is $O\left( \tau _{2}^{\max} BN_{\rm T}L_{\rm RIS} \right) $, where $\tau _{1}^{\max}$ and $\tau _{2}^{\max}$ are the corresponding iteration numbers. Combined with the complexity of solving the SDP problem~\eqref{ABOP-trace} discussed in \textbf{Algorithm~\ref{Optimal Algorithm}}, the total complexity of \textbf{Algorithm~\ref{overall low comeplexity Algorithm}} is $O\left( t_{\max}\left( 2N_{\text{T}}^{6}+\tau _{1}^{\max}\left( L_{\text{RIS}}+1 \right) ^6+ \right. \right. 
 $ $\left. \left. \tau _{2}^{\max}QN_{\text{T}}L_{\text{RIS}} \right) \right) 
 $. It is easy to verify that $\left( \tau _{1}^{\max}\left( L_{\rm RIS}+1 \right) ^6+  \tau _{2}^{\max}QN_{\rm T}L_{\rm RIS} \right)$ is usually much less than $2^{2Q\left( L_{\rm RIS}^{2}+L_{\rm RIS} \right) +1}$, especially for large $L_{\rm RIS}$. Therefore, the proposed low-complexity algorithm is much more efficient compared to \textbf{Algorithm~\ref{Optimal Algorithm}}.
\section{Numerical Results}
In this section, numerical results are provided in order to evaluate the performance of the proposed algorithms. In the three-dimensional (3D) coordinates, the BS and the RIS are located at coordinates $(0~\rm m, 10~\rm m, 0~\rm m)$ and $(80~\rm m, 10~\rm m, 0~\rm m)$, respectively. The coordinates of the NOMA-strong user and the NOMA-weak user are set to $(40~\rm m, 0~\rm m, 0~\rm m)$ and $(80~\rm m, 0~\rm m, 0~\rm m)$, respectively. The distance-dependent path loss is modeled as $P\left( d \right) =\rho \left( d \right) ^{-\alpha}$, where \emph{d} is the link distance, $\alpha$ is the path loss exponent, $\rho=-30~\rm dB$ is the path loss at the reference distance of 1 m. In particular, the path loss exponents for the BS to the NOMA-strong user link and the NOMA-weak user are set to be 3.5 and 4 to distinguish the two users' channel conditions, respectively. As the RIS is deployed close to the NOMA-weak user and far away from the NOMA-strong user, the corresponding path loss exponents are set to be 2.2 and 3.5. respectively. The path loss exponent of the BS-RIS link is  2.2~\cite{fu2019reconfigurable,mu2019exploiting}.
To model the small-scale fading for all channnels involved, we adopt Rician fading, which is given by $f_{\rm Rician}=\sqrt{\frac{\kappa}{1+\kappa}}f_{\rm Rician}^{\text{LoS}}+\sqrt{\frac{1}{1+\kappa}}f_{\rm Rician}^{\text{NLoS}}$, where $\kappa$ is the Rician factor, ${{{f}}_{\rm Rician}^{\rm Los}}$ and ${{{f}}_{\rm Rician}^{\rm NLoS}}$ are the line-of-signt (LoS) component and non-LoS (NLoS) component, respectively. We set the Rician factor $\kappa =2$ for the BS-RIS link and the RIS to the NOMA-weak user link, respectively, and $\kappa =0$ for other communication links. We assume that the operating frequency is $2.5$ GHz and the bandwidth of each channel is 15~kHz, the minimum QoS requirement of the NOMA-strong uer is  ${R_s^{\rm min}}=1~ \rm bit/s/Hz$ and the noise power is ${\sigma ^2} =- 90~\rm dBm$.
\subsection{Convergence of the Proposed Algorithms}
 \vspace{-0.3cm}
Fig.~\ref{Converge} illustrates the convergence behavior of the proposed algorithms versus the iteration number. The proposed AOBO and LCAOBS algorithms converge in about 3 to 4 iterations. In addition, The LCAOBS algorithm can achieve performance close to that achieved by the AOBO algorithm. Though some performance loss is incurred by the LCAOBS algorithm, the complexity of the LCAOBS algorithm is much lower than that of the
AOBO algorithm.
\subsection{Impact of RIS} 
\subsubsection{Total transmit power versus the resolution bits of RIS phase shifts}
\begin{figure}[H]
	\setlength{\belowcaptionskip}{-0.7cm}   
	\centering 
	\begin{minipage}[t]{0.47\textwidth} 
		\centering 
		\includegraphics[scale=0.5]{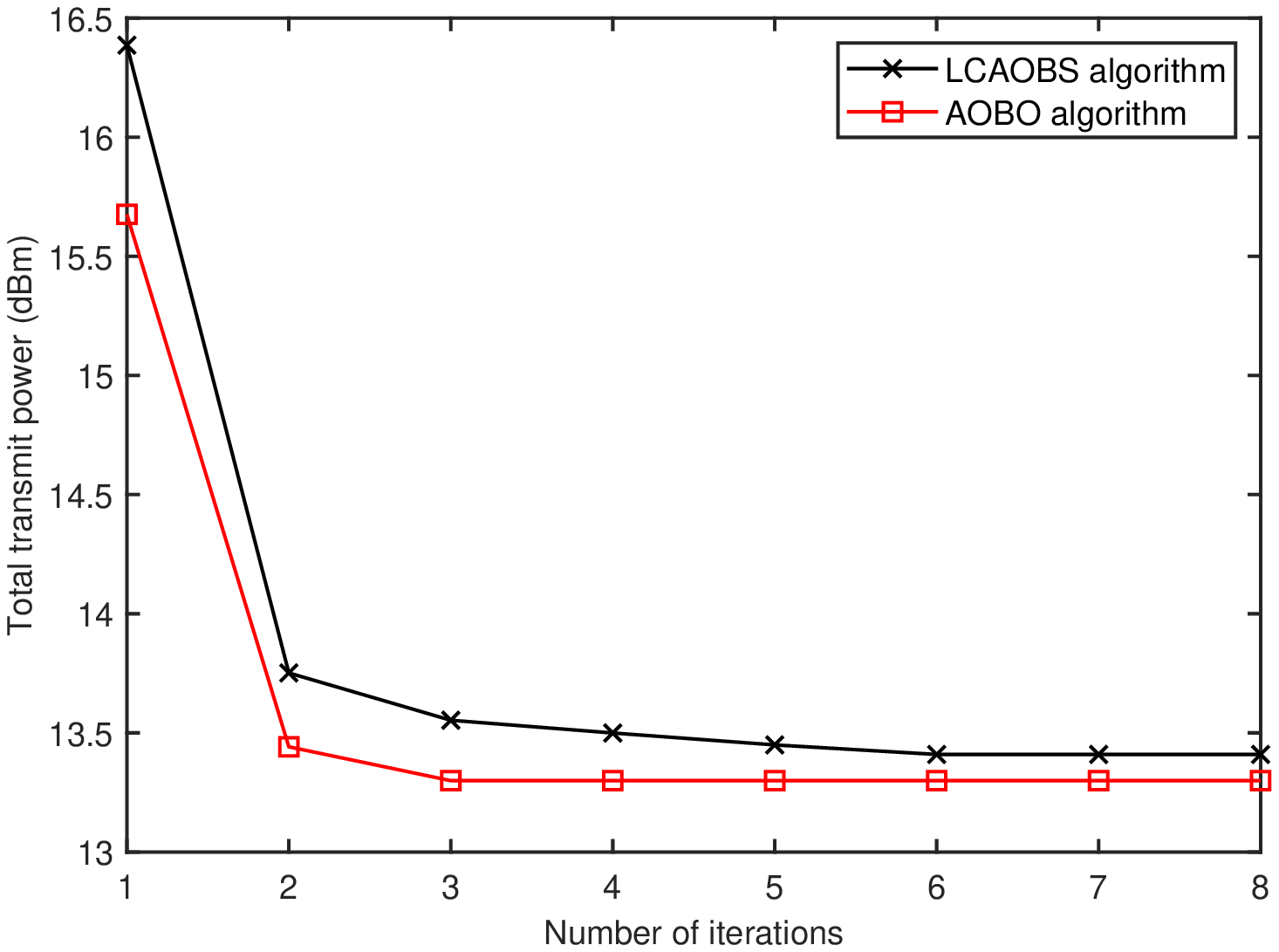}
		\caption{Total transmit power versus the number of iterations, $L_{\rm RIS}=20$, ${R_w^{\rm min}}=2~ \rm bit/s/Hz$, $B=5$}
		\label{Converge}
	\end{minipage}
	\begin{minipage}[t]{0.47\textwidth} 
		\centering 
		\includegraphics[scale=0.5]{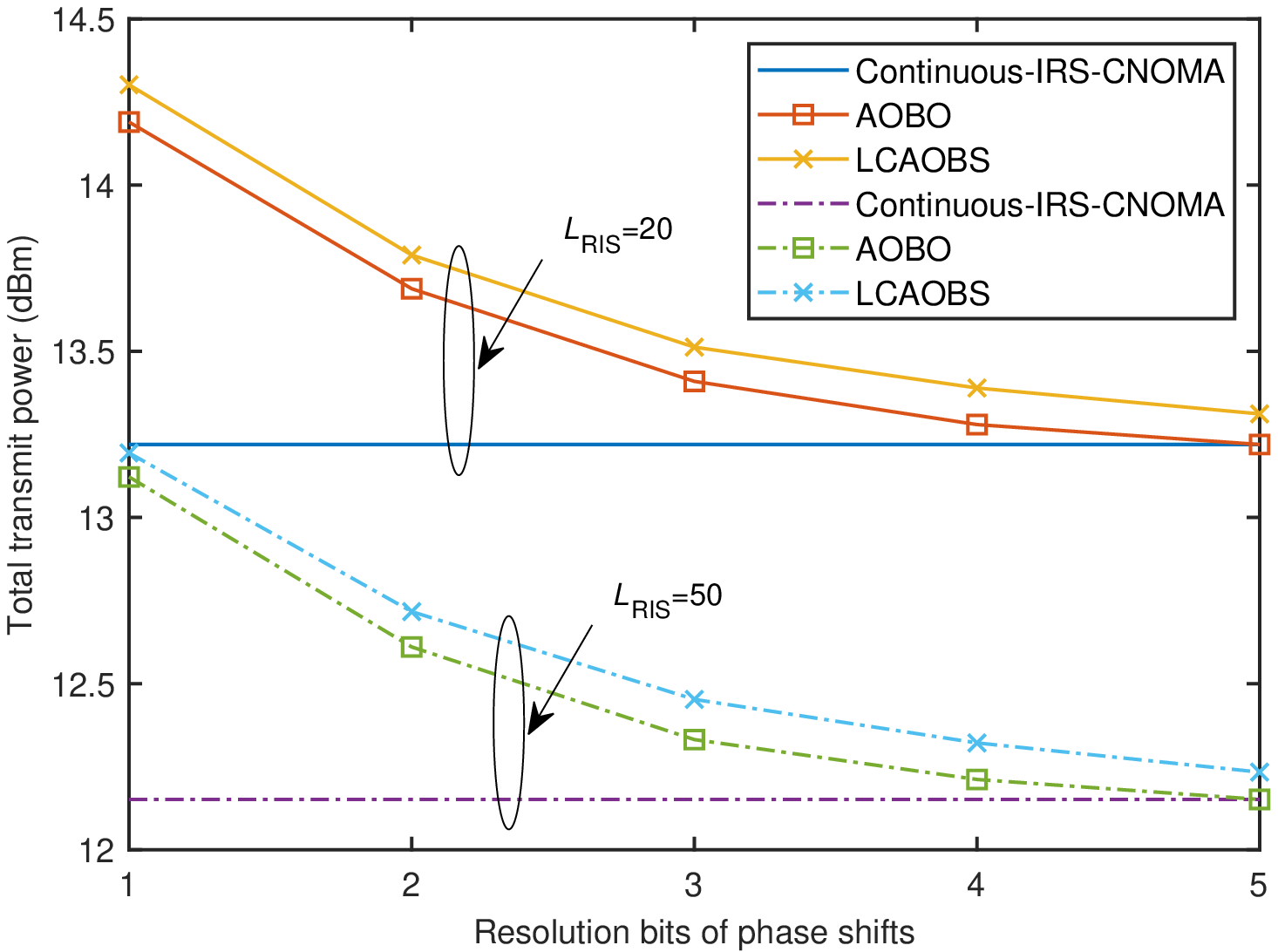}
		\caption{Total transmit power versus the number of phase shifts resolution bits, ${R_w^{\rm min}}=2~ \rm bit/s/Hz$}
		\label{Power_bit}
	\end{minipage}
\end{figure}

Fig.~\ref{Power_bit} depicts the impact of the resolution bits of RIS phase shifts resolution bits on the total transmit power. We compare the proposed algorithms with the Conti-RIS-CNOMA algorithm, which solves problem~\eqref{OOP1} with continuous phase shifts, i.e, $\theta _{m}^{\left( 1 \right)}, \theta _{m}^{\left( 2 \right)} \in \left[ 0,2\pi \right] $, $m=1,2,\cdots ,L_{\text{RIS}}$. It is observed that the total transmit power gap between the continuous and discrete phase shifts gradually decreases as the resolution bits $B$ increases because a larger value  $B$ allows a better adjustment on the RIS phase shifts. However, the implementation complexity increases
in practice with a higher number of resolution bits, especially for the AOBO algorithm. In addition, the proposed LCAOBS algorithm performs close to the AOBO algorithm under different $L_{\rm RIS}$ with low complexity. Based on this observation, we only evaluate the performance of the LCAOBS algorithm in the following subsections because of the high complexity of the AOBO algorithm.
\subsubsection{Total transmit power versus the number of RIS reflecting elements}
\begin{figure}[H]
 \setlength{\belowcaptionskip}{-0.6cm}   
	\centering 
	\begin{minipage}[t]{0.47\textwidth} 
		\centering 
		\includegraphics[scale=0.5]{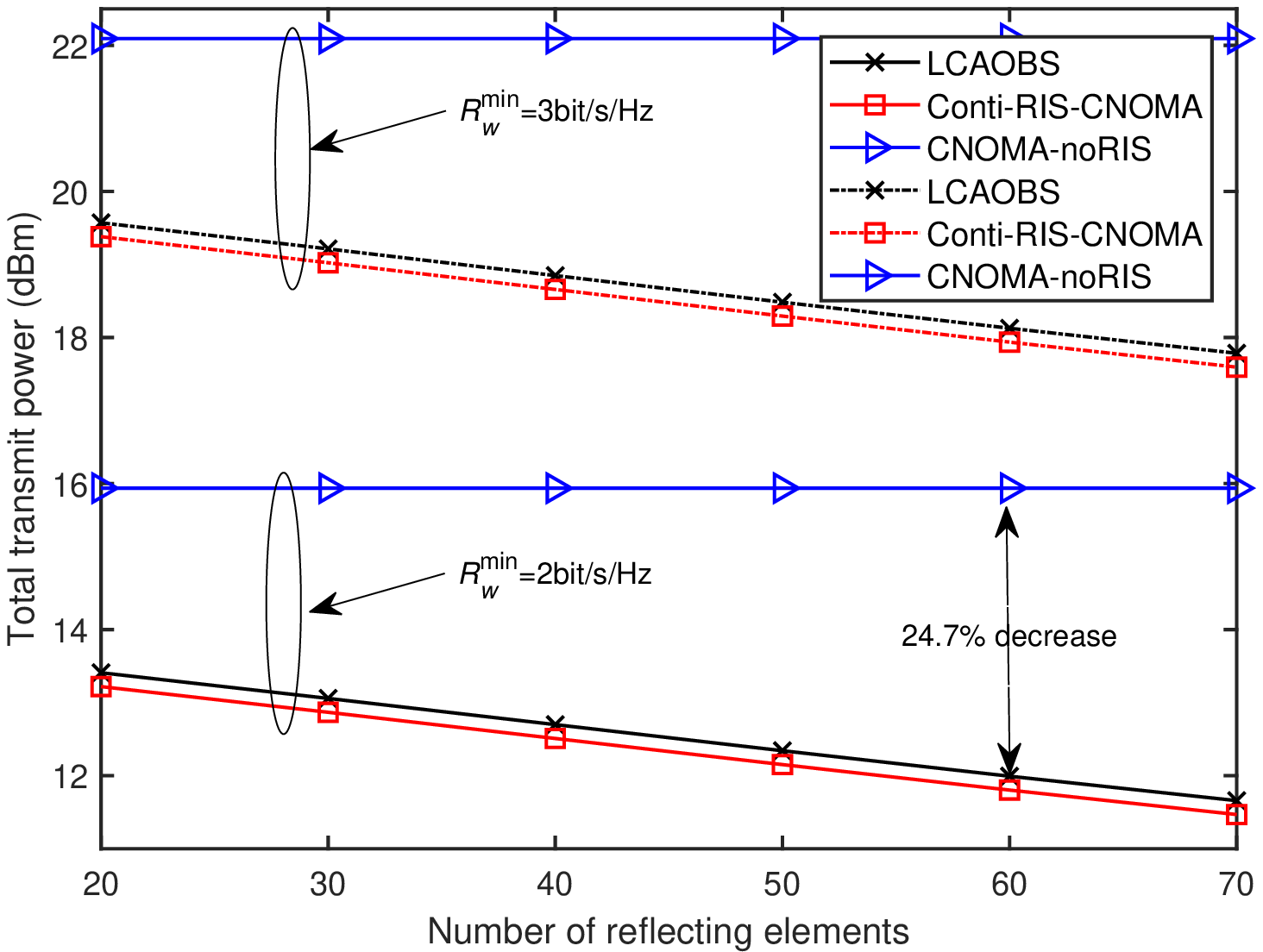}
		\caption{Total transmit power versus the number of reflecting elements, $B=5$}
		\label{Power_LRIS}
	\end{minipage}
	\begin{minipage}[t]{0.47\textwidth} 
		\centering 
		\includegraphics[scale=0.5]{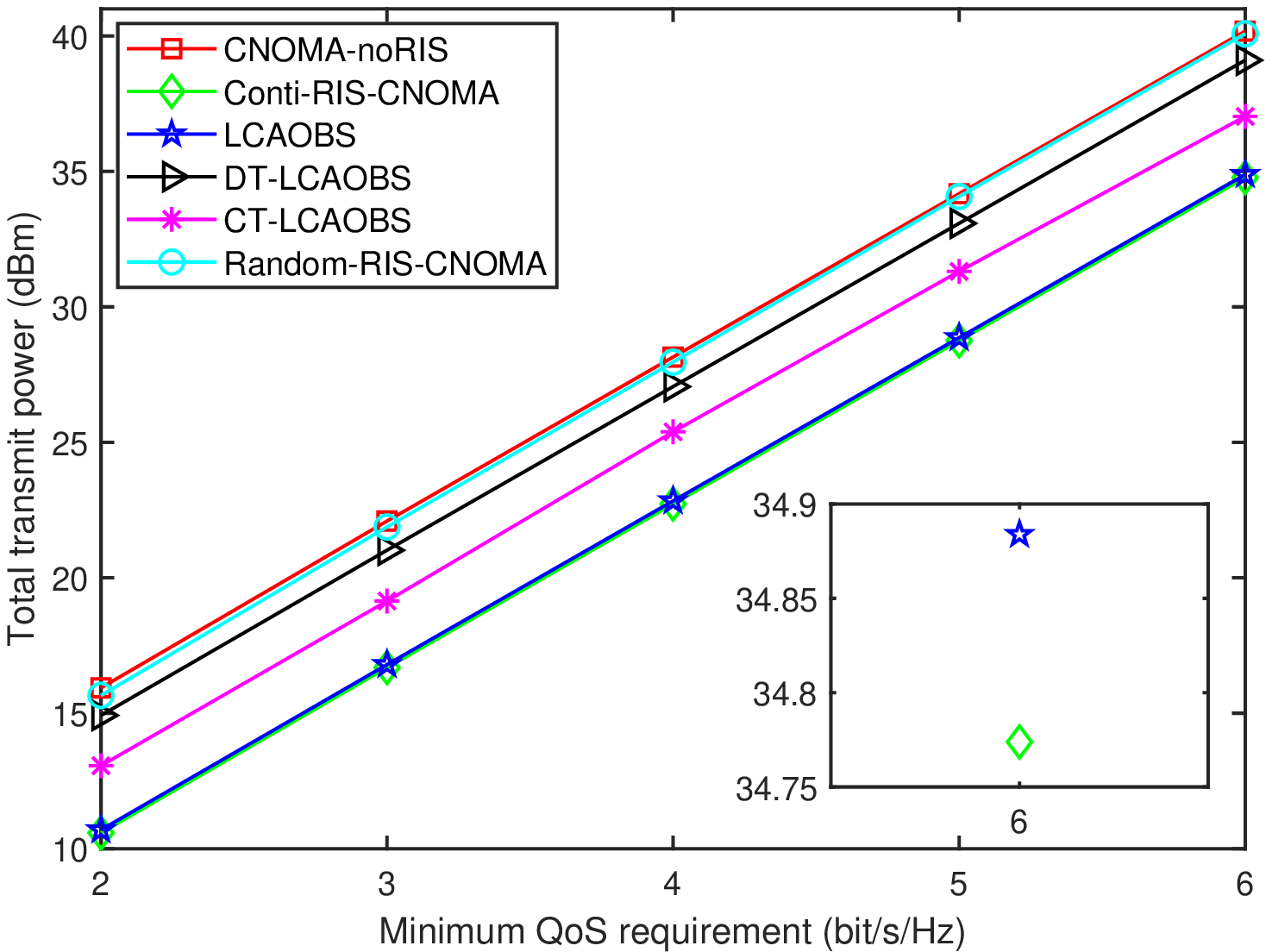}
		\caption{Total transmit power versus the minimum QoS requirement, $L_{\rm RIS}=100$, $B=5$}
		\label{Power_Rmin}
	\end{minipage}
\end{figure}

Fig.\ref{Power_LRIS} shows the total transmit power versus the number of RIS reflecting elements $L_{\rm RIS}$. It is first observed that the total transmit power achieved by the RIS-based algorithms decreases as $L_{\rm RIS}$ increases because a larger number of RIS reflecting elements leads to a higher passive array gains. Second, we observe that the Conti-RIS-CNOMA algorithm has the best performance and the proposed LCAOBS algorithm is capable of closely approaching the Conti-RIS-CNOMA algorithm's performance. Third, the algorithms designed for the RIS-CNOMA system significantly outperforms the CNOMA-noRIS algorithm for the CNOMA system without RIS, which reveals that the application of RIS to the CNOMA system can further improve the power efficiency.
\subsection{Performance Comparison for Different Methods} 
In order to validate the effectiveness of our proposed algorithms, three benchmark schemes are considered, namely, Random-RIS-CNOMA, DT-LCAOBS, CT-LCAOBS. For the Random-RIS-CNOMA algorithm, the DT-stage and CT-stage RIS phase shifts are selected randomly. For the DT-LCAOBS algorithm, we only optimize the DT-stage RIS phase shifts, while the CT-stage RIS phase shifts are selected randomly. In the CT-LCAOBS algorithm, only the CT-stage RIS phase shifts are optimized, while the DT-stage RIS phase shifts are selected randomly. In Fig.~\ref{Power_Rmin}, we present the total transmit power versus the minimum QoS requirement $R_w^{\rm min}$. First, it is observed that the total transmit power of all schemes increases as $R_w^{\rm min}$ becomes more stringent. Second, benefiting from the enhanced combined-channel strength from both the BS and the RIS, the RIS assisted algorithms can decrease the total transmit power compared with the CNOMA-noRIS algorithm for the conventional CNOMA system without RIS. Third, the Conti-RIS-CNOMA an LCAOBS algorithms have the best performance, and the performance gap between the two algorithms is negligible. Finally, a considerable performance loss is observed from the Random-RIS-CNOMA, DT-LCAOBS and CT-LCAOBS algorithms, which conforms the importance of the joint optimization over the DT-stage and the CT-stage RIS phase shifts.
 \subsection{Impacts of RIS Location and User Location}
 Here, we investigate the impacts of the RIS' location and users' locations on the total transmit power and transmit-relaying power. The double coordinates system (DCS) is used to depict the above impacts. In particular, the y-axis in the right-hand side of the DCS is the total transmit power and the y-axis in the left-hand side of the DCS is the transmit-relaying power. 
 
\begin{figure}[H]
	\setlength{\abovecaptionskip}{-0.1cm}
	\setlength{\belowcaptionskip}{-0.5cm}   
	\centering 
	\begin{minipage}[t]{0.47\textwidth} 
		\centering 
		\includegraphics[scale=0.5]{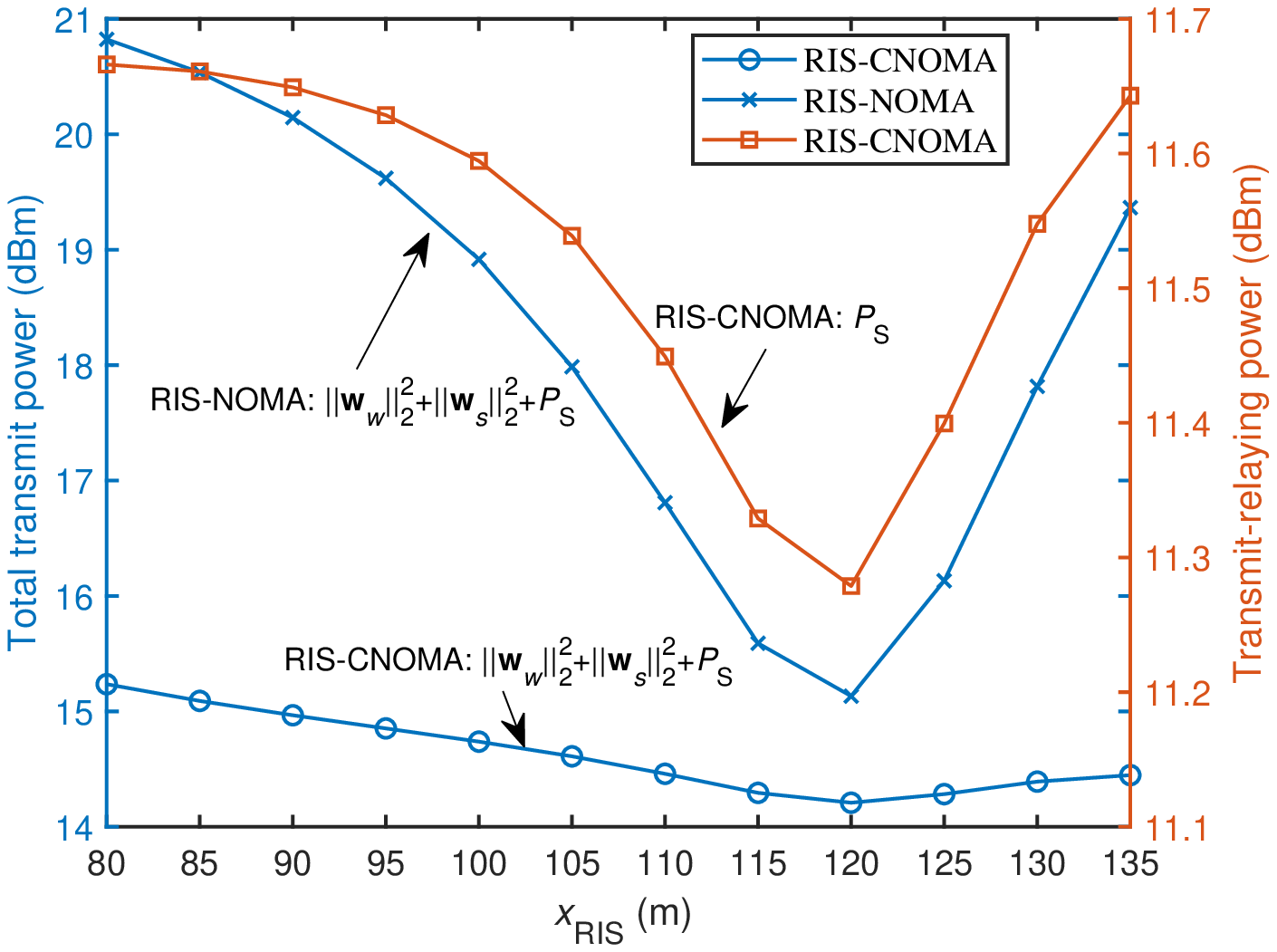}
		\caption{The impact of the RIS' location, $L_{\rm RIS}=20$, ${R_w^{\rm min}}=2~ \rm bit/s/Hz$, $B=5$}
		\label{Power_RIS_location}
	\end{minipage}
	\begin{minipage}[t]{0.47\textwidth} 
		\centering 
		\includegraphics[scale=0.5]{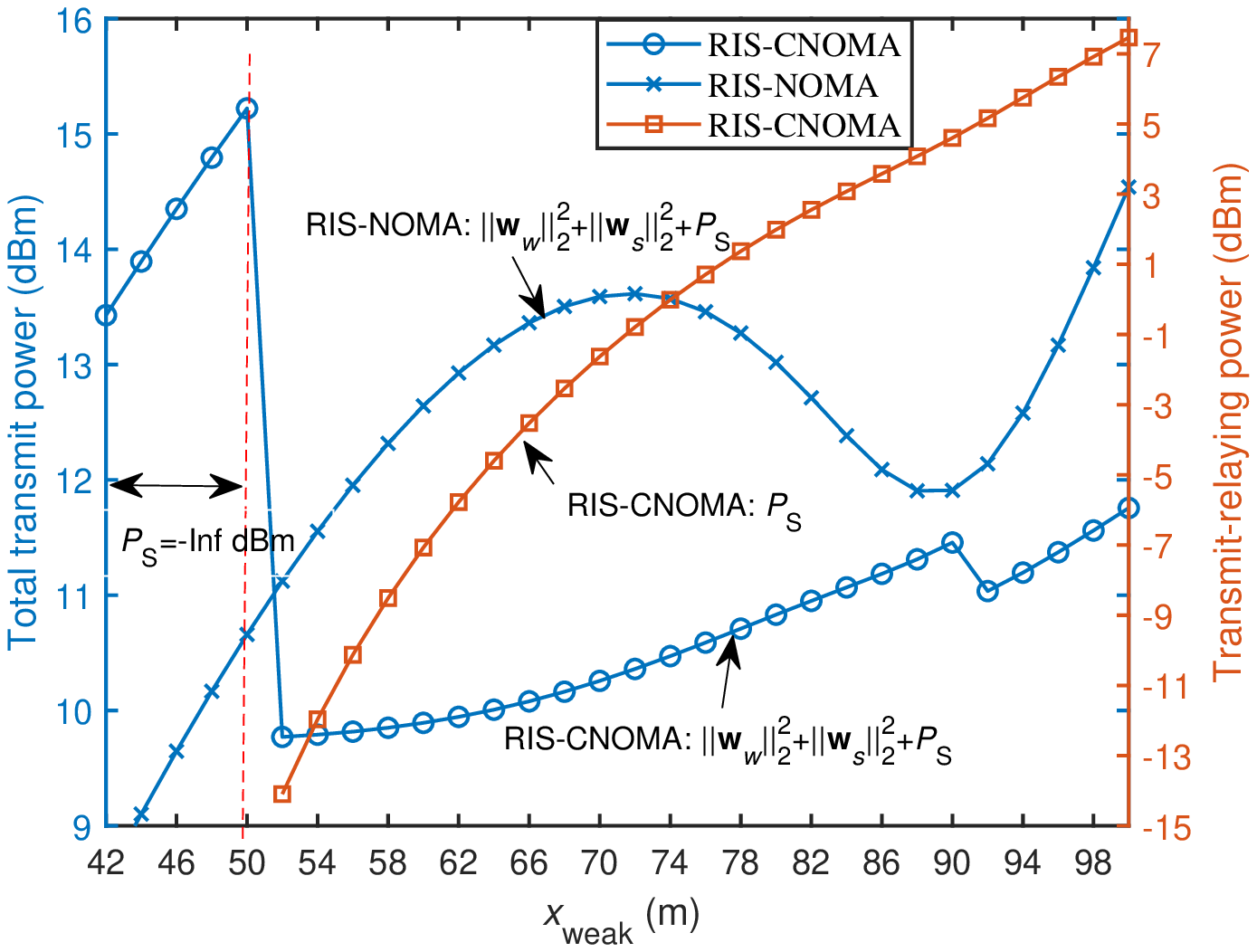}
		\caption{The impact of the NOMA-weak user's location, $L_{\rm RIS}=20$, ${R_w^{\rm min}}=2~ \rm bit/s/Hz$, $B=5$}
		\label{Power_user_location}
	\end{minipage}
\end{figure}

 Fig.~\ref{Power_RIS_location} depicts the total transmit power and transmit-relaying power versus the location of the RIS. Without loss of generality, we set the coordinates of the NOMA-weak user and the RIS as $(120~\rm m, 0~\rm m, 0~\rm m)$ and $(x_{\rm RIS}~\rm m, 10~\rm m, 0~\rm m)$, respectively, where $80 ~{\rm m}\leqslant x_{\text{RIS}}\leqslant 135 ~{\rm m}$. To simplify the analysis, we ignore the small-scale fading
effects. It is observed that the total transmit power and transmit-relaying power achieve their minimum values at $x_{\rm RIS}=120~\rm m$. Because the involved channel gains increase when the RIS gets close to the NOMA-weak user and decreases when it moves away. Furthermore, we observe that the proposed RIS-CNOMA system outperforms the RIS-NOMA system. Because cooperative NOMA can achieve an improved diversity gain for the NOMA-weak user.
In Fig.~\ref{Power_user_location}, we investigate the total transmit power and transmit-relaying power versus the location of NOMA-weak user. The coordinates of the RIS and NOMA-weak user are set to be $(90~\rm m, 10~\rm m, 0~\rm m)$ and $(x_{\rm weak}~\rm m, 0~\rm m, 0~\rm m)$, respectively, where $42 ~{\rm m}\leqslant x_{\text{weak}}\leqslant 100 ~{\rm m}$. It is observed that when $42~{\rm m}\leqslant x_{\rm weak}\leqslant 50~{\rm m}$, the RIS-CNOMA system performs worse than the RIS-NOMA system. Furthermore, the transmit-relaying power $P_{\rm S}=0~{\rm W} ( i.e.,P_{\rm S}= -\rm Inf~ dBm)$ $\footnote{-\rm Inf denotes negative infinity.}$. Beacuse the NOMA-weak user is deployed closer to the BS, thus the minimum QoS requirement of user $w$ can be easily satisfied in the DT stage and the cooperation in the CT stage is not needed. Therefore, the transmit-relaying power $P_{\rm S}=0~\rm W$, which is consistent with \textbf{Remark~\ref{optimal power reamrk}}. Since only the first DT stage is used for signal transmission, the RIS-CNOMA system performs worse than the RIS-NOMA system. Additionally, when $x_{\rm weak}\geqslant  52~{\rm m}$, the RIS-CNOMA system significantly outperforms the RIS-NOMA system. Finally, the RIS-CNOMA and the RIS-NOMA systems achieve their local minimum total transmit power near the RIS and then the total transmit power of both systems increases with the NOMA-weak user moving away from the RIS.
 \section{Conclusions}
In this paper, the RIS assisted CNOMA system has been investigated. The total transmit power was minimized by jointly optimizing the active beamforming vectors, transmit-relaying power and RIS phase shifts. To solve the formulated non-convex problem, an alternating optimization based optimal algorithm was proposed, where the optimal solutions were achieved in each alternation. Due to its prohibitive computational complexity, the above optimal algorithm serves as a benchmark and we proposed a low-complexity algorithm to strike a
trade-off between the performance and complexity. Simulation results showed that the proposed algorithms outperformed the benchmarks and the proposed suboptimal algorithm was capable of achieving near-optimal performance. In particular, compared with conventional CNOMA system, the proposed RIS-CNOMA system can achieve 24.7$\%$ of transmit power savings under 5 resolution bits. Furthermore, when the number of resolution bits is larger than or equal to 5, the performance gap between the discrete phase shifts with 5 resolution bits and the continuous phase shifts was negligible. Additionally, by carefully deploying the RIS' location, the performance gains of the RIS-CNOMA and RIS-NOMA systems can be enhanced. 
In this work, only a half-duplex relay mode was considered for RIS assisted CNOMA systems. The RIS assisted hybrid half/full-duplex CNOMA system is also an important topic to investigate in further work.
\section*{Appendix~A: Proof of Proposition~\ref{Integer expression}} \label{Proof Proposition 1}
\renewcommand{\theequation}{A.\arabic{equation}}
\setcounter{equation}{0}
Since $\textbf{A}$ is a Hemit matrix, we have:
\begin{equation}\label{vAv+Ab}
\begin{array}{l}
\mathbf{v}_{\left( 1 \right)}^{H}\mathbf{Av}_{\left( 1 \right)}+2\mathcal{R}\left( \mathbf{v}_{\left( 1 \right)}^{H}\mathbf{b} \right) 
\\
=2\sum_{m=1}^{L_{\text{RIS}}-1}{\sum_{n=m+1}^{L_{\text{RIS}}}{\mathcal{R}\left\{ \left[ \mathbf{A} \right] _{m,n}e^{j\left( \theta _{m}^{\left( 1 \right)}-\theta _{n}^{\left( 1 \right)} \right)} \right\}}}+2\mathcal{R}\left( \left[ b \right] _me^{j\theta _{m}^{\left( 1 \right)}} \right) +\sum_{m=1}^{L_{\text{RIS}}}{\left[ \mathbf{A} \right] _{m,m}}
\\
=2\sum_{m=1}^{L_{\text{RIS}}-1}{\sum_{n=m+1}^{L_{\text{RIS}}}{\mathcal{R}\left\{ \left[ \mathbf{A} \right] _{m,n}\left( \cos \left( \theta _{m}^{\left( 1 \right)}-\theta _{n}^{\left( 1 \right)} \right) +j\sin \left( \theta _{m}^{\left( 1 \right)}-\theta _{n}^{\left( 1 \right)} \right) \right) \right\}}}
\\
\ \  \  +2\mathcal{R}\left( \left[ b \right] _m\left( \cos \theta _{m}^{\left( 1 \right)}+j\sin \theta _{m}^{\left( 1 \right)} \right) \right) +\sum_{m=1}^{L_{\text{RIS}}}{\left[ \mathbf{A} \right] _{m,m}}
\end{array}.
\end{equation} 

The above function is not linear with respect to $ \theta _m^{\left( 1 \right)},  m=1,2,\cdots ,L_{\text{RIS}}$. However, it can be further transformed into a linear function by exploiting the discrete property of phase shifts  $\left\{ \theta_m^{\left( 1 \right)} \right\} $.
With the definitions of  $\boldsymbol{\delta }_{m}^{\left( 1 \right)}$, $\widetilde{\boldsymbol{\delta }}_{m,n}^{\left( 1 \right)} $, $\mathbf{v}_{\cos}$ and $\mathbf{v}_{\sin}$, we have
\begin{equation}\label{cos sin}
\cos \theta _m^{\left( 1 \right)}=\mathbf{v}_{\cos}^{T}\boldsymbol{\delta }_{m}^{\left( 1 \right)},\sin \theta_m^{\left( 1 \right)}=\mathbf{v}_{\sin}^{T}\boldsymbol{\delta }_{m}^{\left( 1 \right)},
\end{equation}
\begin{equation}\label{cos sin1}
\cos \left( \theta_m^{\left( 1 \right)}-\theta _n^{\left( 1 \right)} \right) =\mathbf{v}_{\cos}^{T}\widetilde{\boldsymbol{\delta }}_{m,n}^{\left( 1 \right)},\sin \left( \theta _m^{\left( 1 \right)}-\theta _n^{\left( 1 \right)} \right) =\mathbf{v}_{\sin}^{T}\widetilde{\boldsymbol{\delta }}_{m,n}^{\left( 1 \right)}.
\end{equation}
 
Substituting~\eqref{cos sin} and \eqref{cos sin1} into~\eqref{vAv+Ab}, we have
 \begin{equation}\label{vAv+Ab simple}
  \begin{array}{l}
	\mathbf{v}_{\left( 1 \right)}^{H}\mathbf{Av}_{\left( 1 \right)}+2\mathcal{R}\left( \mathbf{v}_{\left( 1 \right)}^{H}\mathbf{b} \right) \\
  =2\sum_{m=1}^{L_{\text{RIS}}-1}{\sum_{n=m+1}^{L_{\text{RIS}}}{\mathcal{R}\left\{ \left[ \textbf{A} \right] _{m,n}\left( \left( \textbf{v}_{\cos}^{T}+j\textbf{v}_{\sin}^{T} \right) \widetilde{\boldsymbol{\delta }}_{m,n} \right) \right\}}}\\
\ \  \    +2\mathcal{R}\left( \left[ b \right] _m\left( \textbf{v}_{\cos}^{T}+j\textbf{v}_{\sin}^{T} \right) \boldsymbol{\delta }_m \right) +\sum_{m=1}^{L_{\text{RIS}}}{\left[ \textbf{A} \right] _{m,m}}\\
  \end{array}.
 \end{equation} 
 
It is easy to see that function $\mathbf{v}_{\left( 1 \right)}^{H}\mathbf{Av}_{\left( 1 \right)}+2\mathcal{R}\left( \mathbf{v}_{\left( 1 \right)}^{H}\mathbf{b} \right)$ is linear with respect to $\boldsymbol{\delta }_{m}^{\left( 1 \right)}
$ and $\widetilde{\boldsymbol{\delta }}_{m,n}^{\left( 1 \right)} $. Thus, the proof of \textbf{Proposition~\ref{Integer expression}} is completed.
 \vspace{-0.6cm}
\section*{Appendix~B: Proof of Proposition~\ref{Phase1 algorithm converge} } \label{Proof Proposition XX}
\renewcommand{\theequation}{B.\arabic{equation}}
\setcounter{equation}{0} 
	We denote by $\mathcal{G}\left( \overline{\mathbf{V}}_{\left( 1 \right)}^{\left( \tau_1 \right)} \right)$ the objective value of problem~\eqref{POP trace} for a feasible solution $\overline{\mathbf{V}}_{\left( 1 \right)}^{\left( \tau_1 \right)}
$. Then, we have:
\begin{align}\label{SDP convergence}
\begin{split}
\mathcal{G}\left( \overline{\mathbf{V}}_{\left( 1 \right)}^{\left( \tau_1 +1 \right)} \right) 
&=\mathcal{G}\left( {\rm Tr}\left( \overline{\mathbf{V}}_{\left( 1 \right)}^{\left( \tau_1 +1 \right)} \right) -\varepsilon _{\max}\left( \overline{\mathbf{V}}_{\left( 1 \right)}^{\left( \tau +1 \right)} \right) \right)  
\\
&\overset{\left( a \right)}{\leqslant}\mathcal{G}\left\{ {\rm Tr}\left( \overline{\mathbf{V}}_{\left( 1 \right)}^{\left( \tau_1 +1 \right)} \right) -\varepsilon _{\max}\left( \overline{\mathbf{V}}_{\left( 1 \right)}^{\left( \tau_1 \right)} \right) -\left[ {\rm Tr}\left( \overline{\mathbf{v}}_{\text{EV}}^{\left( \tau_1 \right)}\left( \overline{\mathbf{v}}_{\text{EV}}^{\left( \tau_1 \right)} \right) ^H\left( \overline{\mathbf{V}}_{\left( 1 \right)}^{\left( \tau_1 +1 \right)}-\overline{\mathbf{V}}_{\left( 1 \right)}^{\left( \tau_1 \right)} \right) \right) \right] \right\} 
\\
&\overset{\left( b \right)}{\leqslant}\left\{ {\rm Tr}\left( \overline{\mathbf{V}}_{\left( 1 \right)}^{\left( \tau_1 \right)} \right) -\varepsilon _{\max}\left( \overline{\mathbf{V}}_{\left( 1 \right)}^{\left( \tau_1 \right)} \right) \right\}   =\mathcal{G}\left( \overline{\mathbf{V}}_{\left( 1 \right)}^{\left( \tau_1 \right)} \right) 
\end{split},
\end{align}
where (a) comes from the inequality~\eqref{maxeigen inequality}, (b) holds because $\overline{\mathbf{V}}_{\left( 1 \right)}^{\left( \tau_1 +1 \right)}$ is the optimal solution of problem~\eqref{POP trace3}, while $\overline{\mathbf{V}}_{\left( 1 \right)}^{\left( \tau_1  \right)}$ is feasible to problem~\eqref{POP trace3}. 
 
The inequality in~\eqref{SDP convergence} indicates that the objective value of problem~\eqref{POP trace} is always non-increasing after each iteration. On the other hand, since the objective is lower-bounded, the proposed \textbf{Algorithm~\ref{Low Phase Shifts1 Algorithm}} is guaranteed to converge.
\section*{Appendix~C: Proof of Proposition~\ref{Phase2 algorithm converge} }

\label{Proof Proposition XX}

\renewcommand{\theequation}{C.\arabic{equation}}

\setcounter{equation}{0} 

Let $\mathcal{F}\left( \theta _{1}^{\left( 2,\tau _2 \right)},\theta _{2}^{\left( 2,\tau _2 \right)},\cdots ,\theta _{L_{\text{RIS}}}^{\left( 2,\tau _2 \right)} \right) $ denote the objective value of problem~\eqref{PS3 low} in the $\tau _2$-th iteration, where $\theta _{l}^{\left( 2,\tau _2 \right)}$ is the $\tau _2$-th iteration solution, $l=1,2,\cdots ,L_{\rm RIS}$. According to \textbf{Algorithm~\ref{Low Phase Shifts2 Algorithm}}, we have
\begin{equation}\label{Phase 2 converge}
\begin{split}
\mathcal{F}\left( \theta _{1}^{\left( 2,\tau _2 \right)},\theta _{2}^{\left( 2,\tau _2 \right)},\cdots ,\theta _{L_{\text{RIS}}}^{\left( 2,\tau _2 \right)} \right) 
  & \overset{\left( a \right)}{\leqslant}\mathcal{F}\left( \theta _{1}^{\left( 2,\tau _2 +1\right)},\theta _{2}^{\left( 2,\tau _2 \right)},\cdots ,\theta _{L_{\text{RIS}}}^{\left( 2,\tau _2 \right)} \right) 
\\
& \vdots 
\\
& \leqslant \mathcal{F}\left( \theta _{1}^{\left( 2,\tau _2 +1\right)},\theta _{2}^{\left( 2,\tau _2 +1\right)},\cdots ,\theta _{L_{\text{RIS}}}^{\left( 2,\tau _2 +1 \right)} \right) 
\end{split},
\end{equation} 
where (a) holds since for fixed $\left\{ \theta _{2}^{\left( 2,\tau _2 \right)},\cdots ,\theta _{l}^{\left( 2,\tau _2 \right)},\cdots ,\theta _{L_{\text{RIS}}}^{\left( 2,\tau _2 \right)} \right\} $, $\theta _{1}^{\left( 2,\tau _2+1 \right)}$ is the optimal solution to problem~\eqref{PS2 low rewritten}. Furthermore, because the phase shifts are optimized in the order from $l=1$ to $l=L_{\rm RIS}$ , $\theta _{l}^{\left( 2,\tau _2+1 \right)}$ is the optimal solution with given $\left\{ \theta _{1}^{\left( 2,\tau _2+1 \right)},\cdots ,\theta _{l-1}^{\left( 2,\tau _2+1 \right)},\theta _{l+1}^{\left( 2,\tau _2 \right)},\cdots ,\theta _{L_{\text{\rm RIS}}}^{\left( 2,\tau _2 \right)} \right\} $. Therefore, the other inequalities in~\eqref{Phase 2 converge} also hold. 

\eqref{Phase 2 converge} indicates that the objective function's value of problem~\eqref{PS2 low rewritten} is non-decreasing after each iteration. In addition, since $\mathbf{v}_{\left( 2 \right)}^{H}\mathbf{Z}_w\mathbf{v}_{\left( 2 \right)}\leqslant L_{\rm RIS}\varepsilon _{\max}\left( \mathbf{Z}_w \right) 
$ and $\mathcal{R}\left( \mathbf{v}_{\left( 2 \right)}^{H}\mathbf{u}_w \right) \leqslant \sum_{m=1}^{L_{\rm RIS}}{\lvert \left[ \mathbf{u}_w \right] _m \rvert}$, we have
\begin{equation}\label{Phase 2 upper bound}
\mathbf{v}_{\left( 2 \right)}^{H}\mathbf{Z}_w\mathbf{v}_{\left( 2 \right)}+2\mathcal{R}\left( \mathbf{v}_{\left( 2 \right)}^{H}\mathbf{u}_w \right) +\lvert h_{s,w} \rvert^2\leqslant L_{\rm RIS}\varepsilon _{\max}\left( \mathbf{Z}_w \right) +2\sum_{m=1}^{L_{\rm RIS}}{\lvert \left[ \mathbf{u}_w \right] _m \rvert}+\lvert h_{s,w} \rvert^2,
\end{equation}
which means that the objective value of problem~\eqref{PS2 low rewritten} is upper-bounded by a finite value.

Based on the above analysis, we conclude that the proposed algorithm is guaranteed to converge.

 \vspace{-0.5cm}
 \bibliographystyle{IEEEtran}
 \bibliography{zjkbib}

\end{document}